\documentclass[%
 aip,
 amsmath,amssymb,
reprint, 
]{revtex4-1}

\usepackage{graphicx}
\usepackage{dcolumn}
\usepackage{bm}

\usepackage[utf8]{inputenc}
\usepackage[T1]{fontenc}
\usepackage{mathptmx}
\usepackage{etoolbox}

\makeatletter
\def\@email#1#2{%
 \endgroup
 \patchcmd{\titleblock@produce}
  {\frontmatter@RRAPformat}
  {\frontmatter@RRAPformat{\produce@RRAP{*#1\href{mailto:#2}{#2}}}\frontmatter@RRAPformat}
  {}{}
}%
\makeatother
\begin{document}

\preprint{AIP/123-QED}

\title{Dopant ionization and efficiency of ion and electron ejection from helium nanodroplets}
\author{Jakob D. Asmussen}
\affiliation{Department of Physics and Astronomy,
	Aarhus University, Denmark}
 \author{Ltaief Ben Ltaief}
\affiliation{Department of Physics and Astronomy,
	Aarhus University, Denmark}
 \author{Keshav Sishodia}
\affiliation{Quantum Center for Diamond and Emergent Materials (QuCenDiEM) and Department of Physics, Indian Institute of Technology Madras, Chennai 600036, India}
  \author{Abdul R. Abid}
\affiliation{Department of Physics and Astronomy, Aarhus University, Denmark}
 \author{Bj\"orn Bastian}
\affiliation{Department of Physics and Astronomy, Aarhus University, Denmark}
 \author{Sivarama Krishnan}
\affiliation{Quantum Center for Diamond and Emergent Materials (QuCenDiEM) and Department of Physics, Indian Institute of Technology Madras, Chennai 600036, India}
\author{Henrik B. Pedersen}
\affiliation{Department of Physics and Astronomy,
	Aarhus University, Denmark}
 \author{Marcel Mudrich}
\email{mudrich@phys.au.dk}
\affiliation{Department of Physics and Astronomy,
	Aarhus University, Denmark}

\date{\today}

\begin{abstract}
Photoionization spectroscopy and mass spectrometry of doped helium (He) nanodroplets rely on the ability to efficiently detect ions and/or electrons. Using a commercial quadrupole mass spectrometer and a photoelectron-photoion coincidence (PEPICO) spectrometer, we systematically measure yields of ions and electrons created in pure and doped He nanodroplets in a wide size range and in two ionization regimes -- direct ionization and secondary ionization after resonant photoexcitation of the droplets. For two different types of dopants (oxygen molecules, O$_2$, and lithium atoms, Li), we infer the optimal droplet size to maximize the yield of ejected ions. When dopants are ionized by charge-transfer to photoionized He nanodroplets, the highest yield of O$_2$ and Li ions is detected for a mean size of $\sim5\times10^4$ He atoms per nanodroplet. When dopants are Penning ionized via photoexcitation of the He droplets, the highest yield of O$_2$ and Li ions is detected for $\sim10^3$ and $\sim10^5$ He atoms per droplet, respectively. At optimum droplet sizes, the detection efficiency of dopant ions in proportion to the number of primary photoabsorption events is up to 20\,\% for charge-transfer ionization of O$_2$ and 2\,\% for Li, whereas for Penning ionization it is 1\,\% for O$_2$ and 4\,\% for Li.
Our results are instrumental in determining optimal conditions for mass spectrometric studies and photoionization spectroscopy of molecules and complexes isolated in He nanodroplets.
\end{abstract}

\maketitle

\section{Introduction}
Helium nanodroplets (HNDs) are cold (0.38~K) superfluid large clusters of weakly-bound He atoms.~\cite{barranco2006helium} Due to their transparency from infrared to ultraviolet light and their ability to pick-up and efficiently cool molecules and complexes (‘dopants’) embedded in a chemically inert environment, HNDs are often considered as `the ideal spectroscopic matrix'.~\cite{toennies2004superfluid} However, when dopants are photoionized, the favorable conditions presented by HNDs for spectroscopy of neutral species are generally no longer well fulfilled; strong polarization forces acting between ions and the helium tend to induce significant line shifts and nuclear dynamics.~\cite{mudrich2014photoionisaton,gonzalez2020solvation,asmussen2023electron} Still, valuable information about molecular spectra~\cite{mudrich2004formation,giese2011homo,loginov2008new,smolarek2010ir}, the formation of metal clusters~\cite{tiggesbaumker2007formation,theisen2011rb,kazak2019photoelectron,messner2018spectroscopy} and fast chemical dynamics~\cite{braun2004imaging,giese2012formation,kautsch2015photoinduced,von2017imaging,thaler2020ultrafast,stadlhofer2022dimer,asmussen2022time} among others have been achieved through photoionization spectroscopy. 

Large HNDs are currently arousing interest due to their ability to form unusual nanostructures.~\cite{haberfehlner2015formation,messner2020shell,alic2023diamondoid} Aside from surface deposition~\cite{schiffmann2020helium,martini2021splashing} and coherent diffraction imaging~\cite{gomez2014shapes,langbehn2022diffraction} used as new techniques for probing HND-aggregated nanostructures, mass spectrometry remains an important diagnostic technique.~\cite{thaler2015synthesis,mauracher2018cold} However, ions tend to remain attached to large HNDs~\cite{atkins1959ions,muller2009alkali,theisen2010forming} which prevents them from efficiently being detected. Accordingly, mass spectra of dopants ionized in HNDs display only that fraction of ions which are ejected from the HNDs whereas a large fraction of ions may remain undetected. Therefore a better understanding of the ejection mechanisms of both ions and electrons out of HNDs of different sizes is needed to interpret photoionization and photoelectron spectra and to optimize conditions that rely on the detection of ions and/or electrons, \textit{e.~g.} mass spectrometric studies and photoionization spectroscopy. 

An optimal HND size is obtained as a compromise between two effects: On the one hand, the pick-up cross-section for dopants scales with the geometrical cross section of the HND; a HND with radius $R$ containing $N\approx 4\pi n_\mathrm{He}R^3/3$ He atoms picks up a number of dopants $k\propto R^2\propto N^{2/3}$.~\cite{harms1998density} Here, $n_\mathrm{He}\approx 0.022~$\AA$^{-3}$ is the number density of superfluid He. On the other hand, ions are efficiently solvated in larger HNDs and therefore elude detection unless the ions are produced in an excited state~\cite{smolarek2010ir,zhang2012communication,loginov2012spectroscopy,von2015dynamics} or at high kinetic energy~\cite{pickering2018femtosecond}. Likewise, in large HNDs electrons are trapped in void bubbles; these bubbles migrate to the HND surface where electrons remain bound in metastable states before tunneling out into the vacuum.~\cite{farnik1998differences,rosenblit2006electron} Therefore, we expect that the efficiency of detecting ions and/or electrons from dopant ionization is maximum at some intermediate size of the HNDs.

In this work, we report measurements of the yields of ejected ions and electrons using the photoelectron-photoion coincidence (PEPICO) technique for HNDs of variable size and with a dopant residing either on the surface or in the bulk of the HND. Most dopants are immersed in the HND interior; only alkali metal atoms reside at the HND surface in a dimple due to Pauli repulsion between He and the diffuse valence electron of the alkali atom.~\cite{barranco2006helium} In this study, we use lithium (Li) as surface-bound dopant and oxygen molecules (O$_2$) as a dopant immersed in the bulk of the HND. HNDs are resonantly photoexcited by extreme ultraviolet (XUV) synchrotron radiation at a photon energy $h\nu = 21.6~$eV and photoionized at $h\nu = 26~$eV. 

Using PEPICO detection, we have previously investigated various energy and charge-transfer (CT) processes initiated by photoexcitation or photoionization of the HND.~\cite{buchta2013extreme,buchta2013charge,laforge2016enhanced,shcherbinin2017interatomic,shcherbinin2019inelastic,ben2019charge,laforge2019highly,ltaief2020electron} PEPICO detection, including its variants where multiple electrons and ions are detected in coincidence, is a powerful tool for deciphering ionization and fragmentation processes in great detail~\cite{jarvis1999high,ueda2005molecular,bastian2022new}; however, it relies on the efficient detection of both electrons and ions. 
Here, we infer the optimal HND size for PEPICO detection of ejected ions and electrons by analyzing in detail the electron-ion-coincidence mass spectra recorded for pure and doped HNDs, and we quantify the contribution of uncondensed He atoms in the HND beam expansion as function of expansion temperature. 
We study the HND-size dependence on coincidence electron-ion yields for two different ionization mechanisms. At $h\nu=21.6$~eV, the HND is resonantly excited into an absorption band mainly associated with the $1s2p\,^1$P state of atomic He. The excited He atom undergoes ultrafast relaxation into the lowest metastable atomic states~\cite{mudrich2020ultrafast,laforge2022relaxation} from which it further relaxes to the ground state by ionizing a dopant through an interatomic Coulombic decay (ICD) process~\cite{averbukh2004mechanism,kuleff2010ultrafast}, traditionally called Penning ionization.~\cite{peterka2006photoionization,buchta2013charge,ben2019charge} In the case of multiple excitation of HNDs, ICD can also occur between two or more excited He atoms.~\cite{ovcharenko2014novel,laforge2014collective,laforge2021ultrafast,asmussen2022time,ltaief2023efficient} At higher photon energies $h\nu>24.6~$eV, the HND is directly ionized and the dopant can in turn be ionized through CT processes to the He cation.~\cite{buchta2013charge,ellis2007model} 
As the effective range of ICD or CT ionization processes is limited,~\cite{ben2019charge,ellis2007model} the ion yield decreases with increasing HND size~\cite{buchta2013charge,stadlhofer2022dimer}. Here we report and discuss the optimal HND sizes for the two ionization regimes. 

\section{Experimental setup}
Yields of photoions and electrons were measured using the PEPICO technique applied to pure and doped HNDs at the XENIA (XUV electron-ion spectrometer for nanodroplets in Aarhus) endstation~\cite{bastian2022new} located at the AMOLine of the ASTRID2 synchrotron at Aarhus University, Denmark.~\cite{hertel2011astrid2} With this endstation, we can record velocity map images (VMIs) for all emitted electrons, electrons in coincidence with cations as well as ion time-of-flight mass spectra. Electron and ion yields were background-subtracted using a spinning chopper wheel that periodically blocks the HND beam. For resonant excitation ($h\nu = 21.6$~eV) of the HNDs, a Sn filter was inserted to block higher orders of the undulator radiation; when HNDs were directly ionized ($h\nu = 26.0$~eV), an Al-filter was used to attenuate the beam. 
All spectra are normalized to the photon flux. With the current setup used for photoionization, we cannot detect total yields of ions in a mass-resolved fashion. Instead, we used a commercial quadrupole mass spectrometer (QMS, model Extrel 5221) that ionizes the HND beam by electron impact from a hot filament to detect mass-resolved total ion yields. The impact electron energy was set to $E_e = 70$~eV. 

HNDs were formed by continuous expansion of high-purity He gas at high pressure (30~bar) into vacuum through a cryogenically cooled nozzle of diameter 5~$\mu$m. The HND size was adjusted by varying the temperature of the nozzle from 7 to 32~K resulting in a mean HND radius in the range of $R=1$-$190$~nm corresponding to a mean number size of $\langle N \rangle = 10$-$10^9$. $\langle N \rangle $ was determined by titration measurements~\cite{gomez2011sizes} and by comparison to literature values.~\cite{toennies2004superfluid} HNDs were doped with about one Li atom on average by passing them through a 1~cm long heated to a temperature around 400~$^\circ$C. Doping with O$_2$ was achieved by leaking O$_2$ gas into a doping chamber with a length of 22~cm located between the expansion chamber and the spectrometer. The O$_2$ doping level was adjusted to monomer doping conditions across the different HND sizes.

\section{Results and discussion}
\subsection{Total electron and ion yields from pure helium nanodroplets}
\begin{figure}[t]
    \centering
    \includegraphics[width=0.9\columnwidth]{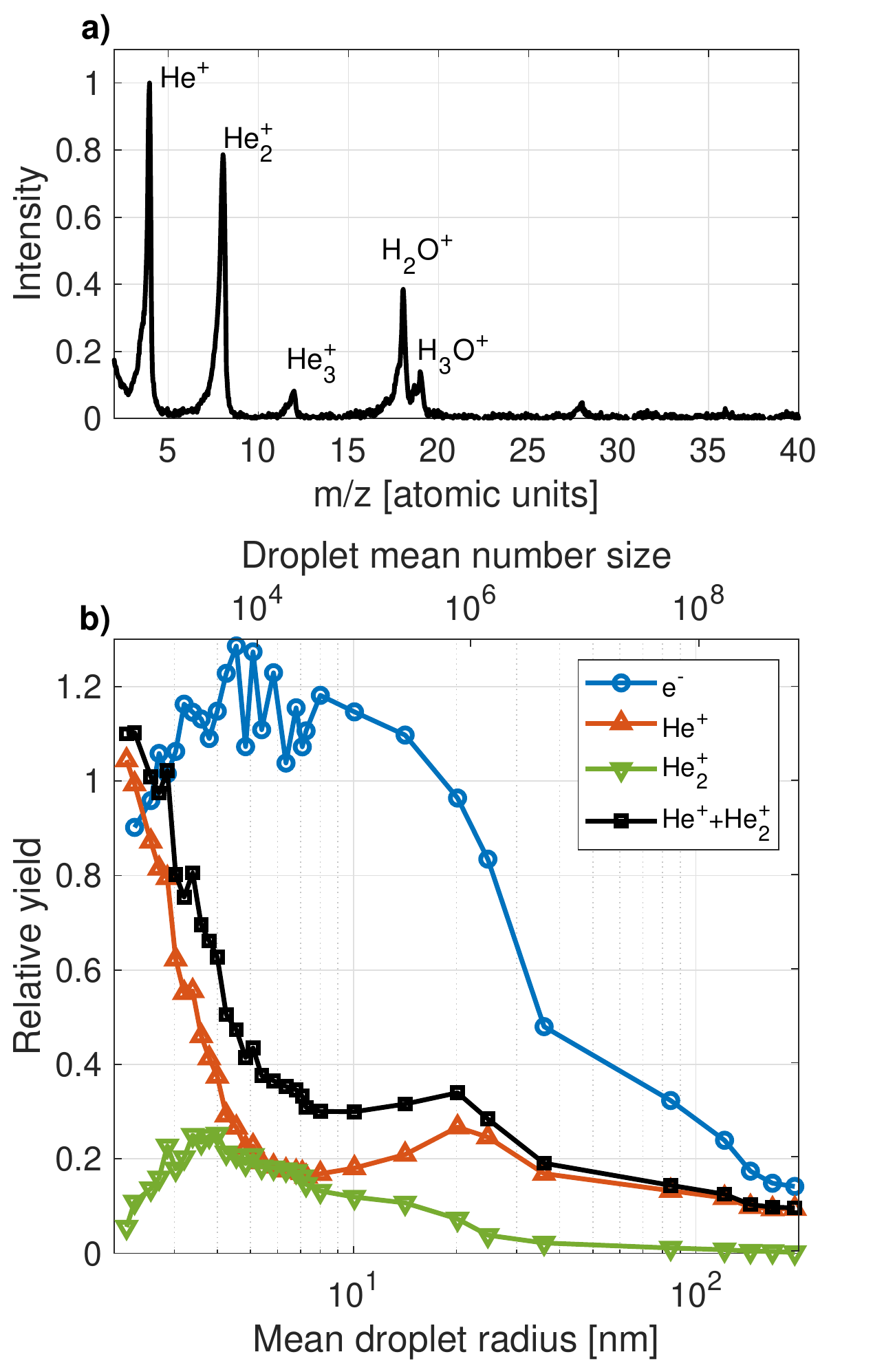}
    \caption{ a) Ion spectrum of pure HNDs recorded by electron impact ionization using the QMS at $E_e = 70$~eV and a droplet size of $\langle N \rangle \sim 10^4$. 
    b) Electron yield relative to the total number of ionization events measured by photoionization ($h\nu = 26$~eV, blue symbols) and relative ion yields measured by electron impact ionization ($E_e = 70$~eV, red, green, and black symbols). 
    }
    \label{fig:total_yield}
\end{figure}
We start by presenting the ion yield measured by electron impact ionization of pure HNDs using the QMS and the detected total electron yield measured by photoionization using the VMI spectrometer. Fig.~\ref{fig:total_yield}~a) shows a typical mass spectrum recorded with the QMS at $E_e = 70$~eV for a mean droplet size $\langle N \rangle \sim 10^4$.  
He$^+$ ions can originate from the atomic component of the beam, from secondary ionization processes that release He$^+$ with non-vanishing kinetic energy~\cite{shcherbinin2017interatomic} and from multiply ionized droplets.
He$_2^+$ and He$_3^+$ ions originate from HNDs; likewise, H$_2$O$^+$ and H$_3$O$^+$ are due to CT ionization of water and water clusters in the droplets formed by pick-up of the residual background gas. 

Electrons can lose most of their kinetic energy by elastic scattering in large HNDs and can be trapped in the HND through electron-ion recombination or solvation by forming a bubble~\cite{asmussen2023electron}. Thus, it is important to determine the emission efficiency of electrons from HNDs as it may be the limiting factor when detecting electron-ion coincidences. However, in this experiment, we do not have a direct measure of the total photoionization or electron impact ionization rates $\Gamma_\mathrm{tot}$ which the detected electron and ion yields should be normalized to in order to account for the increasing He target density, $n_\mathrm{He}$, for lower expansion temperatures. We merely estimate $n_\mathrm{He}$ from the total He flux measured by blocking and unblocking the HND beam and monitoring the difference in pressure in the droplet beam dump chamber, taking into account the temperature-dependent HND speed.~\cite{gomez2011sizes} From $n_\mathrm{He}$ and the estimated flux of photons (VMI) and electrons (QMS), $\Phi_\mathrm{ph,e}$, we estimate $\Gamma_\mathrm{tot}=\Phi_\mathrm{ph,e}\times n_\mathrm{He}\times\ell\times\sigma\propto n_\mathrm{He}$ assuming that the ionization probability per He atom is independent of the HND size. Here, $\ell$ is the length of the interaction region and $\sigma$ is the effective ionization cross section of He atoms by photons (VMI) or electron impact (QMS).  
In this relation, we assume that all electrons and ions escape the HNDs and are detected in the limit of small HNDs, \textit{i.~e.} for sizes $\langle N \rangle < 10^3$ ($R<3$~nm).
Thus, the normalized yields of ions and electrons $\Gamma_\mathrm{ion,e}/\Gamma_\mathrm{tot}\rightarrow 1$ for $\langle N \rangle < 10^3$ ($R<3$~nm). These relative electron and He$^+$ and He$_2^+$ ion yields are depicted in Fig.~\ref{fig:total_yield}~b).
The blue symbols in Fig.~\ref{fig:total_yield}~b) show that all photoelectrons at $h\nu = 26$~eV are detected up to a HND size $\langle N \rangle \sim 10^6$ ($R\sim25$~nm), but electrons are partly trapped in the HNDs at larger sizes. For a HND size $\langle N \rangle > 10^8$ ($R>80$~nm), only 20\,\% of the electrons escape the HNDs. 
The fact that the relative yield seems to exceed 1 around $\langle N \rangle\approx 10^4$ ($R\approx 5$~nm) shows that we slightly underestimate $\Gamma_\mathrm{tot}$ from the He pressure increase.

Unlike the yield of all detected electrons, the yield of He ions measured with the QMS, $\Gamma_\mathrm{ion}/\Gamma_\mathrm{tot}$, rapidly decreases as the HND size increases. Assuming that each HND is only singly ionized, the atomic ion, He$^+$, is not expected to be ejected out of a HND due to its efficient binding to the HND by forming a snowball complex.~\cite{gonzalez2020solvation} However, large HNDs can be multiply ionized owing to their large total cross section resulting in Coulomb explosion of the ions. This likely explains the non-vanishing yield of He$^+$ detected at HND radii $\gtrsim 10$~nm. Furthermore, secondary ionization processes such as ICD of excited ions may further enhance the He$^+$ yield by Coulomb explosion.~\cite{shcherbinin2017interatomic}  

For very small HNDs ($R<1$~nm), the He ion yield mostly consists of He$^+$. As the droplet size increases, the He$^+$ yield drops and the He$_2^+$ yield rises. This is due to a decrease of the contribution of uncondensed He atoms to the He jet and because the efficiency of forming molecular ions increases with HND size in the case of multiple ionization of the HNDs. The total relative He ion yield (sum of He$^+$ and He$_2^+$) drops to 40\,\% at $\langle N \rangle \sim 10^4$ ($R\sim5$~nm), where the yields of the two ion fragments are equal. For even larger HNDs, the ion yield drops to 20\,\%, and the He$^+$ cation becomes again the more abundant ion. 
When HNDs are photoionized near the ionization threshold at low photon flux (low ionization probability per HND and below the threshold for exciting He ions) we would expect the total ion yield to be even smaller. A small local maximum in the He$^+$ yield is measured around $R = 20$~nm ($\langle N \rangle = 2.5 \times10^6$) corresponding to expansion conditions near the critical point.~\cite{toennies2004superfluid} Most likely this is due to break-up of the HNDs near the skimmer where an increased uncondensed contribution of He atoms is generated.  

Overall the comparison between total electron and ion yields reveals that ion solvation by forming stable complexes bound to the HND is a much more efficient process than electron solvation for HNDs of sizes $\langle N \rangle < 10^6$ ($R<30$~nm). Therefore we ascribe any HND-size dependence of the electron-ion coincidence yield to ion solvation and we neglect electron trapping in this size range.

\subsection{Ion yields from pure He droplets measured in coincidence with electrons}

\begin{figure}[t]
    \centering
    \includegraphics[width=1\columnwidth]{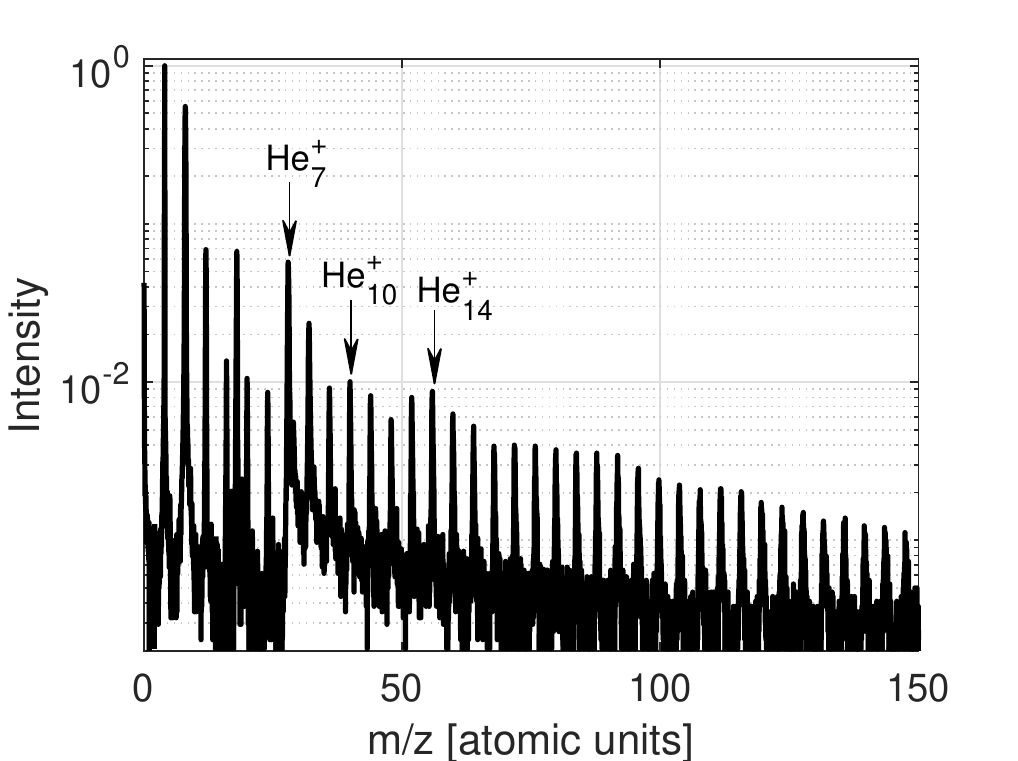}
    \caption{Electron-ion coincidence mass spectrum recorded for $h\nu = 26$~eV and a droplet size of $\langle N \rangle \sim 10^4$. Helium cluster ions He$_n$ with magic numbers $n=7$, $n=10$ and $n=14$ are marked by arrows. }
    \label{fig:mass_pure}
\end{figure}

\begin{figure}[t]
    \centering
    \includegraphics[width=0.8\columnwidth]{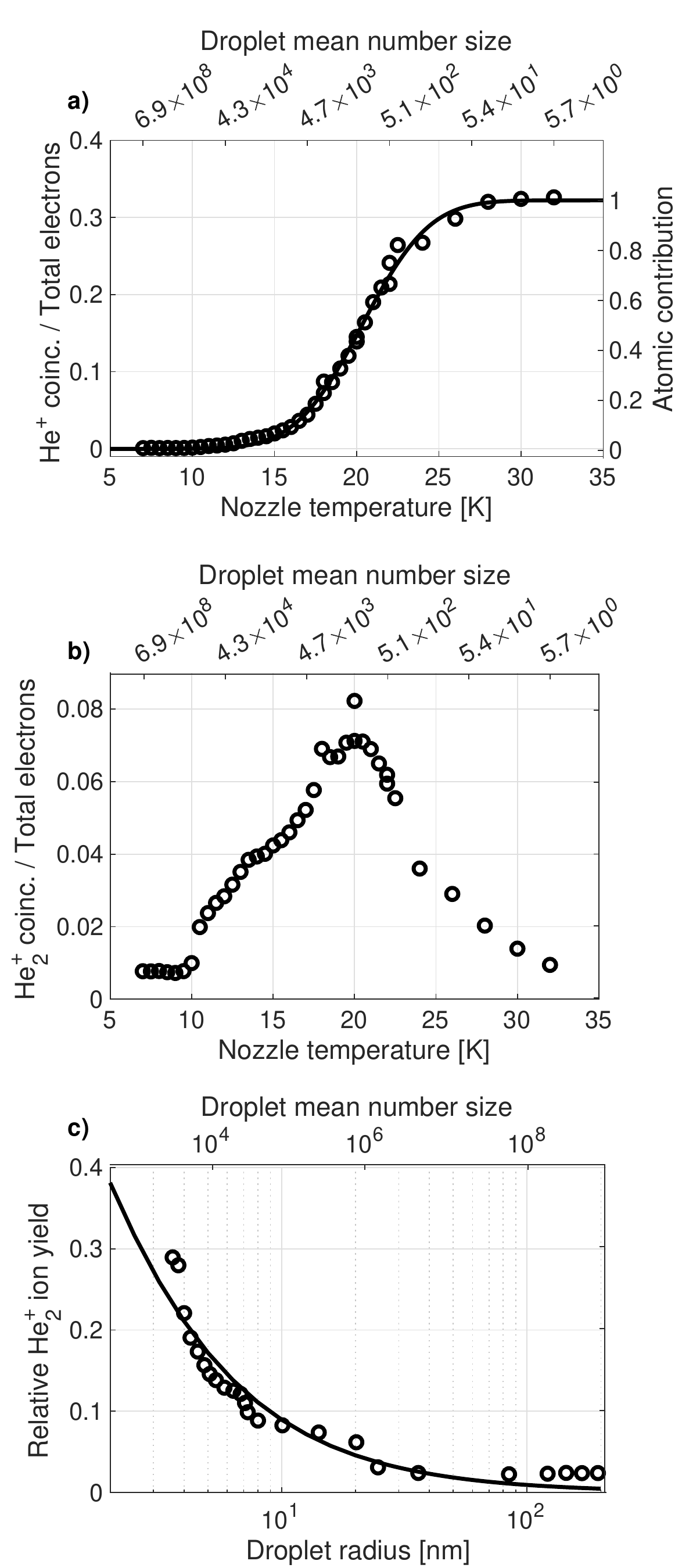}
    \caption{The yield of (a) $e$/He$^+$ coincidences and (b) $e$/He$_2^+$ coincidences relative to the yield of all electrons as function of the nozzle temperature; the upper $x$-axes show the corresponding mean droplet size. The atomic contribution in the He beam is determined from the relative yield of He$^+$. The solid line in (a) shows the results from fitting it with an error function. The relative yield of He$_2^+$-electron coincidences to only HND ionization events (c) is calculated from the droplet contribution of the total electron yield according to Eq.~(\ref{eq:Relative_He2}). The solid line shows the fit according to Eq.~\ref{eq:He2_yield}. 
    The photon energy was 26~eV.}
    \label{fig:pure}
\end{figure}

When photoionizing pure HNDs near the ionization threshold, a series of He cationic fragments are detected in coincidence with photoelectrons (He$_n^+$, $n=1,2,3,\dots$). Fig.~\ref{fig:mass_pure} shows a mass spectrum recorded at $h\nu = 26$~eV for HNDs of size $\langle N \rangle \sim 10^4$. 
Compared to Fig.~\ref{fig:total_yield}~a), larger He ionic fragments ($n\geq4$) can be seen in mass spectra due to the lower noise level achievable with the PEPICO spectrometer. 
The appearance of the so-called ``magic numbers'' ($n=7, \, 10, \, 14$) where He$_n^+$ fragment clusters are more stable matches with reports in the literature.~\cite{stephens1983experimental,callicoatt1998fragmentation,buchenau1990mass} 
Upon ionization of the HNDs, the He$^+$ undergoes charge-hopping from one He atom to another before it localizes by forming a snowball complex in the HND.~\cite{buchenau1991excitation,ellis2007model} Since the photon energy is below the adiabatic double-ionization threshold of HNDs~\cite{shcherbinin2017interatomic}, only a single He$^+$ is formed which remains bound in the droplet. Thus, the detected free He$^+$ ions are exclusively formed from the atomic component in the He nozzle expansion in contrast to impact ionization using the QMS. This is also evident from the photoelectron angular distributions and photoelectron spectra detected in coincidence with He$^+$ (not shown), which are identical to those detected for a beam of free He atoms.~\cite{buchta2013extreme,asmussen2023electron} Thus, we identify the contribution of uncondensed He atoms in the He jet with the relative yield of $e$/He$^+$ coincidences. Fig.~\ref{fig:pure}~a) shows the yield of $e$/He$^+$ coincidences relative to all detected electrons, $\Gamma_{e/\mathrm{He^+}}/\Gamma_e$, as a function of nozzle temperature. We find that it resembles an S-shaped curve showing no effect of the phase transition when the nozzle temperature crosses the critical temperature ($10.5$~K).~\cite{toennies2004superfluid} As the nozzle is cooled further, condensation in the expansion becomes more efficient and the atomic contribution decreases. The temperature dependence is fitted by an error function. 
The value reached in the limit of high expansion temperatures (0.32) thus indicates the relative detection efficiency of the ion detector with respect to the electron detector. We assume equal detection efficiency for all ion masses in the following analysis.~\cite{hellsing1985performance}

Fig.~\ref{fig:pure}~b) shows the yield of $e$/He$_2^+$ coincidences relative to all detected electrons, $\Gamma_{e/\mathrm{He_2^+}}/\Gamma_e$. He$_2^+$ is by far the most abundant cation detected from HNDs ($>$50\,\% of all HND-correlated coincidence ions). Therefore, larger ion fragments from HNDs are neglected in the further analysis. As the HND component in the He jet rapidly increases in the HND size range $5\lesssim N\lesssim 10^3$ whereas the He atomic component decreases, the yield of He$_2^+$ coincidences also rises. For HNDs of sizes $\langle N \rangle > 10^3$, the coincidence yield drops again due to the effect of solvation of ions in the HND. 

By subtracting the atomic contribution determined from the $e$/He$^+$ yield, we can determine the number of detected photoelectrons coming only from the HND component of the He jet. In this way, we obtain the relative He$_2^+$ ion yield from only HND-ionization events,
\begin{equation} \label{eq:Relative_He2}
Y_\mathrm{He_2^+} = \frac{\Gamma_{e/\mathrm{He_2^+}}}{\Gamma_e - \Gamma_{e/\mathrm{He^+}}}. 
\end{equation}
This relative $He_2^+$ yield is shown in Fig.~\ref{fig:pure}~c) as a function of HND size. It drops as the droplet size increases indicating that the $He_2^+$ ions formed in HNDs are efficiently solvated when the HND size increases. To model this behaviour, we assume that cations formed closer to the HND surface are more likely to leave the HND and to be detected. Our recent study of the photoelectron-energy loss due to elastic scattering revealed that all detected ions are formed at a distance $<5$~nm below the HND surface~\cite{asmussen2023electron}. Here, the probability for a He$_2^+$ formed at a distance $r$ from the HND center to escape the HND and to be detected is assumed to be
\begin{equation} \label{eq:He2_prob}
    P_{\mathrm{esc}}(r,R) = e^{-(R-r)/R_{\mathrm{esc}}},
\end{equation}
where $R$ is the HND radius. The HND-size-depended coincidence electron-He$_2^+$ yield is taken as
\begin{equation} \label{eq:He2_yield}
    I^{(\mathrm{He_2^+})}(R) = \frac{4\pi}{V} \int_0^R P_{\mathrm{esc}}(r,R)r^2dr,
\end{equation}
where $V=\frac{4}{3}\pi R^3$ is the volume of the droplet taken as a hard sphere. In Eq.~(\ref{eq:He2_yield}), we neglect the size distribution of the HNDs assuming that all HNDs have the mean size $\langle N\rangle$. Additionally we assume homogeneous illumination of the HNDs. At $h\nu=26$~eV, the penetration depth of photons in He bulk is $67.6$~nm.~\cite{samson2002precision} As the droplet diameter reaches a size larger than the photon penetration depth ($\langle N\rangle>10^6$), the yield of detected He$_2^+$ nearly vanishes meaning that any inhomogeneous illumination effect should not affect the detected ion yield.  

The solid black line in Fig.~\ref{fig:pure}~c) shows the best fit of $Y_\mathrm{He_2^+}$ using Eq.~(\ref{eq:He2_yield}). The fit yields $R_{\mathrm{esc}} = 0.35$~nm showing that only cations formed in the outer-most surface layer of the HND are detected. This value should be treated with caution, though, given the simplicity of the model. Since the detected ions mostly come from the outer surface, a more detailed model should take the radial He density profile of HNDs into account.~\cite{harms1998density,stringari1987systematics} 
He$_2^+$ ions formed near the surface of the HNDs are ejected due to the release of their binding energy ($2.35$~eV).~\cite{callicoatt1998fragmentation} The excess energy from forming He$_2^+$ in the HND bulk is most likely dissipated through boiling off neutral He atoms.~\cite{laimer2019highly} 

The general trend in detecting photoions from pure droplets is that their yield rapidly decreases for increasing HND size. The yield of He$^+$ decreases due to the decreasing atomic fraction of the He jet, whereas the yield of He$_2^+$ decreases because of an increasing contribution of the bulk of the droplets where  He$_2^+$ tend to be solvated. The highest relative yield of droplet-correlated ions is detected for a droplet size of $\langle N \rangle \sim 10^3$.   

\subsection{Yields of dopant ions detected in coincidence with electrons}
\begin{figure*}[t]
    \centering
    \includegraphics[width=1.8\columnwidth]{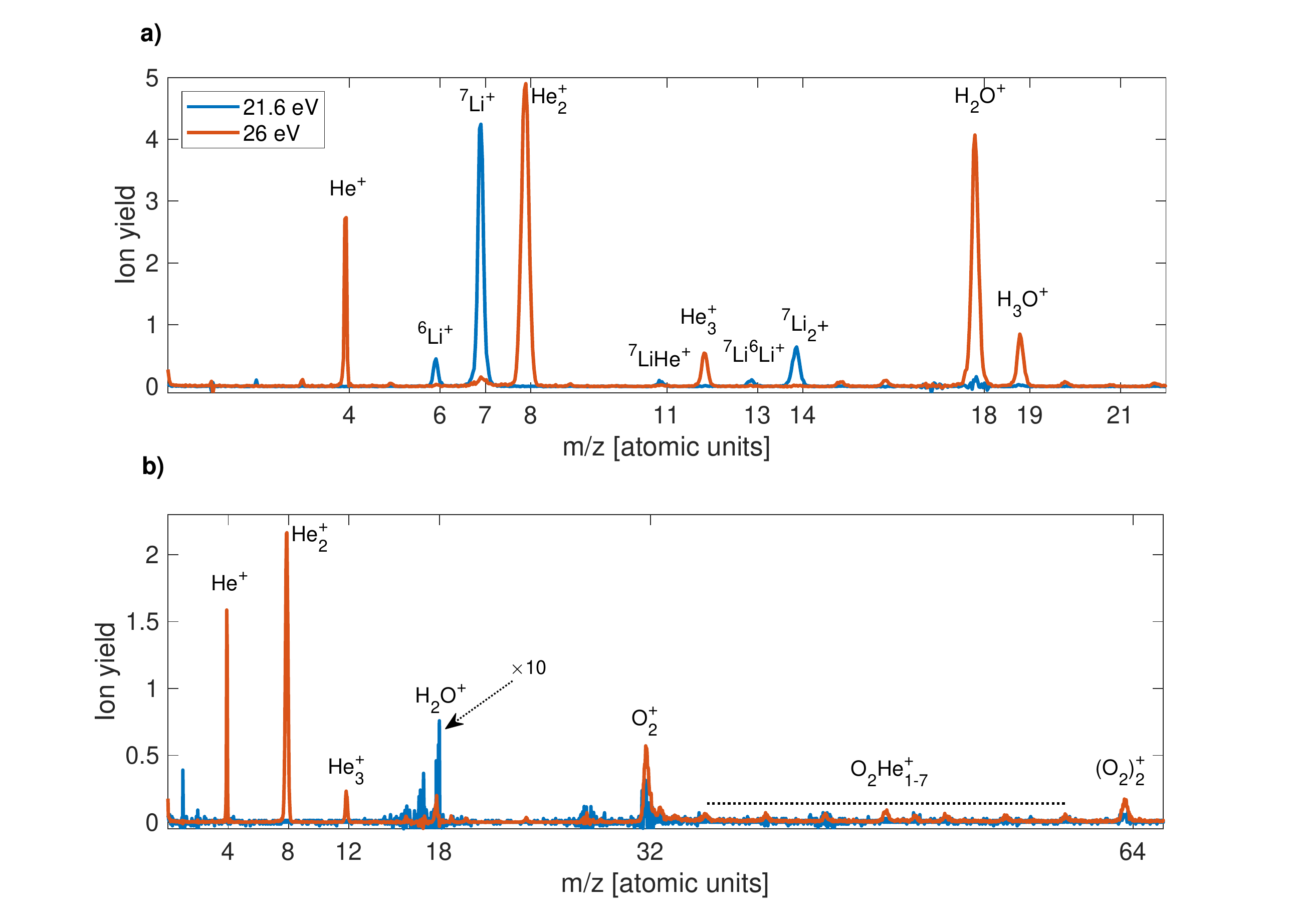}
    \caption{Mass spectra recorded by PEPICO detection when photoionizing HNDs doped with Li (\textbf{a}) and with O$_2$ (\textbf{b}) at $h\nu = 21.6$ and $26$~eV. The average HND size was $\langle N \rangle = 1.6\times10^5$~atoms and the doping strength was $9\times10^{-7}$~mbar$\times$m. All spectra are normalized to the acquisition time and the photon flux. The mass spectra for O$_2$-doped HNDs recorded at $h\nu = 21.6$~eV are scaled with a factor 10 for better visualization.  }
    \label{fig:massspec}
\end{figure*}

More relevant for most photoionization and mass spectrometric studies is the question how efficiently dopant ions are formed and ejected out of HNDs. To address this question, we either resonantly excite or directly photoionize HNDs doped with Li or O$_2$. Fig.~\ref{fig:massspec} shows mass spectra recorded for HNDs with an average size $\langle N \rangle = 1.6\times10^5$~atoms and a doping strength (dopant partial pressure $\times$ doping cell length) $n_l=p_\mathrm{dop}\times l=9\times10^{-7}$~mbar$\times$m for these two dopants. Here $p_\mathrm{dop}$ is the partial pressure of the dopant and $l$ is the length of the doping cell. Higher ion yields of Li dopants [Fig.~\ref{fig:massspec}~a)] are detected for Penning ionization ($h\nu = 21.6$~eV) compared to CT ionization ($h\nu = 26.0$~eV), whereas the opposite is the case for O$_2$ [Fig.~\ref{fig:massspec}~b)]. A similar difference in ionization efficiencies was reported for alkali-doped HNDs in comparison with HNDs doped with heavier rare gas atoms submerged in the HND interior.~\cite{buchta2013charge} Upon excitation, repulsion between the excited He atom and the surrounding He leads to the formation of a bubble around the excited atom and ejection of the atom to the droplet surface~\cite{mudrich2020ultrafast}, thereby enhancing the probability for Penning ionization of species residing at the surface.~\cite{buchta2013charge} On the other hand, a He cation efficiently forms a larger He$_n^+$ snowball complex,~\cite{buchenau1991excitation} which is solvated in the HND. Therefore, CT ionization is more likely to occur for submerged dopant species. This can also be seen from the efficient CT ionization of HNDs doped with water (H$_2$O$^+$) and water clusters (H$_3$O$^+$) from the residual gas which was present when operating the Li-containing heated cell. For both Li-doped and O$_2$-doped HNDs we detect cationic complexes containing He atoms (LiHe$_n^+$, O$_2$He$_n^+$), in particular for CT ionization. Helium cations (He$_n^+$) are only detected for direct photoionization ($h\nu = 26.0$~eV); the photon energy used for resonant excitation, $h\nu = 21.6$~eV, is below the HND ionization energy and no ions should be detected.~\cite{peterka2003photoelectron} However, for large HNDs, $\langle N \rangle > 10^7$, ($R>36$~nm), even at $h\nu = 21.6$~eV we detect He cations due to multiple excitation of the HNDs leading to ICD of pairs of excited He atoms (not shown).~\cite{laforge2021ultrafast,ltaief2023efficient}  


\begin{figure}[t]
    \centering
    \includegraphics[width=0.9\columnwidth]{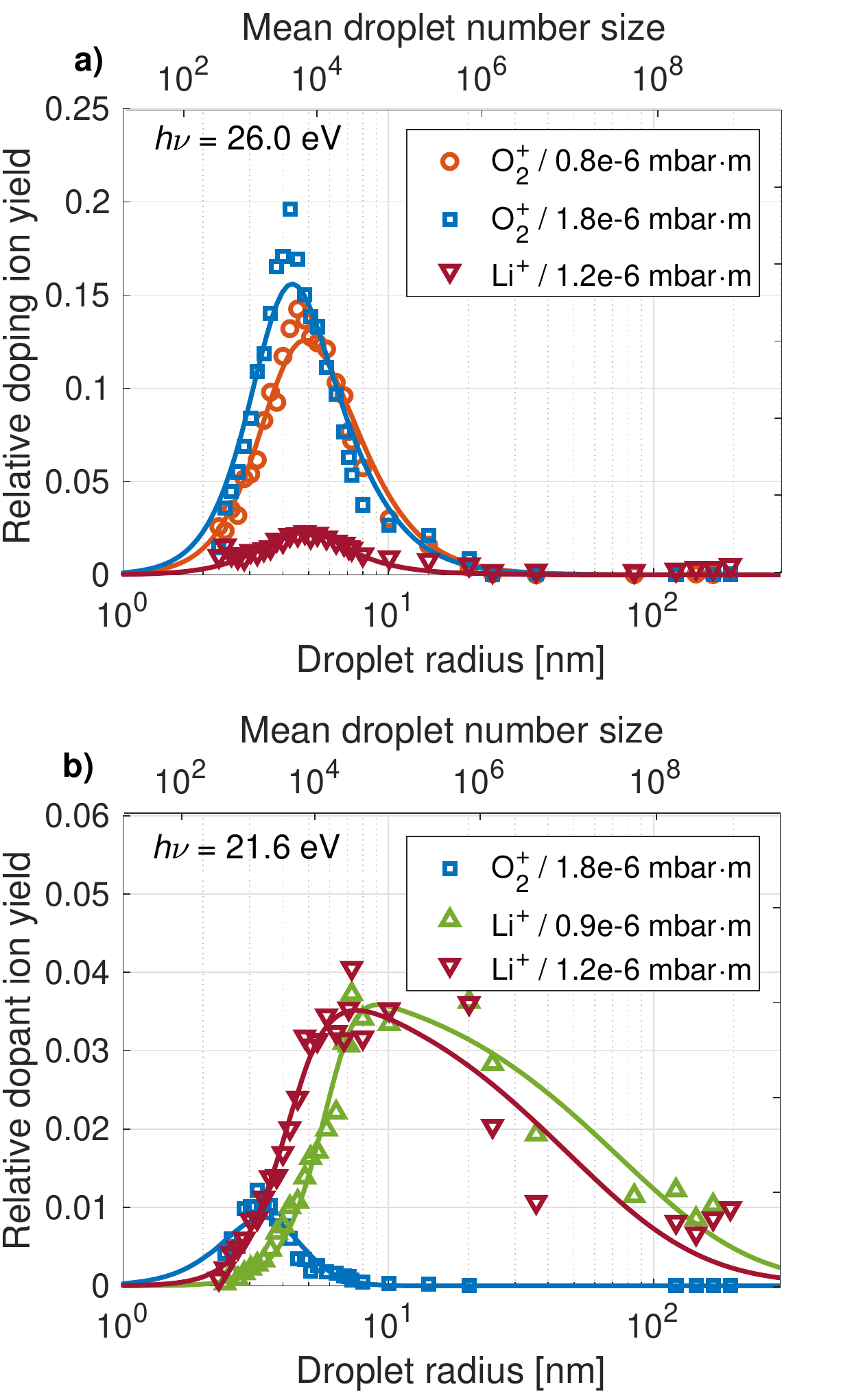}
    \caption{Relative dopant ion yield as function of mean HND size shown for CT ionization (\textbf{a}) according to Eq.~(\ref{eq:Relative_CT}) and Penning ionization (\textbf{b}) according to Eq.~(\ref{eq:Relative_Penning}) for different dopants (O$_2$ and Li) and different doping strengths. The solid lines show fit curves using model functions Eq.~(\ref{eq:CT_O2}), (\ref{eq:CT_Li}), (\ref{eq:P_O2}) and (\ref{eq:P_Li}). } 
    \label{fig:doped}
\end{figure}

Fig.~\ref{fig:doped} shows the yield of electron-dopant ion coincidences relative to all HND-correlated photoabsorption events. In the case of CT ionization [Fig.~\ref{fig:doped}~a)], the relative yield can be written as 
\begin{equation}
    \label{eq:Relative_CT}
Y_\mathrm{Li^+,O_2^+} = \frac{\Gamma_{e/\mathrm{Li^+,O_2^+}}}{\Gamma_e - \Gamma_{e/\mathrm{He^+}}}.
\end{equation}
In the case of Penning ionization [Fig.~\ref{fig:doped}~b)], an electron is only produced in the case when Penning ionization happens. As the photoabsorption rate is not measured directly, we calculate it by scaling to the one at $h\nu = 26$~eV where the photoabsorption rate is directly measured as electron yield. Thus, the relative yield of electron-dopant ion coincidences is given by 
\begin{equation}
    \label{eq:Relative_Penning}
    Y_\mathrm{Li^+,O_2^+} = \frac{\Gamma_{e/\mathrm{Li^+,O_2^+}}}{\Gamma_{\mathrm{He^*}}},
\end{equation}
where $\Gamma_{\mathrm{He^*}}=(\sigma\Phi)_\mathrm{21.6\,eV}/(\sigma\Phi)_\mathrm{26\,eV}\times (\Gamma_e - \Gamma_{e/\mathrm{He^+}})$. Here, $(\sigma\Phi)_\mathrm{21.6\,eV}$ is the absorption cross section multiplied by the photon flux at photon energy $h\nu=21.6~$eV~\cite{buchta2013extreme,samson2002precision}.

Both the yields of $e$/O$_2^+$ coincidences and $e$/Li$^+$ coincidences turn out to feature maxima at intermediate HND sizes. However, their maxima are clearly mutually shifted; $e$/O$_2^+$ coincidences are systematically peaked as smaller HND sizes than $e$/Li$^+$ coincidences. In case of CT ionization [$h\nu = 26$~eV, panel a)], the yield is highest at a mean HND radius $R=4.3$~nm for O$_2^+$ and at $R=4.8$~nm for Li$^+$ coincidence yields. In the case of Penning ionization [$h\nu = 21.6$~eV, panel b)], the maxima are at $R=3.2$~nm and $R=7.2$~nm, respectively. In contrast to the $e$/He$_2^+$-coincidences, the relative dopant ion yield decreases for small HNDs. This is due to the fact that the pick-up cross-section for the dopants depends on the geometrical size of the HND,
\begin{equation} \label{eq:pickup_cs}
    \sigma_{\mathrm{pickup}}(R) = S(R)\sigma_{\mathrm{coll}} = S(R)\pi R^2.
\end{equation}
Here, the cross-section for HND-dopant collisions $\sigma_{\mathrm{coll}}$ is assumed to be the geometrical cross section of the HND. $S$ is the size-depended sticking coefficient determining the likelihood of a collision leading to capture of the dopant. $S=1$ is often assumed for interpretation of doped-HND experiments using HND of sizes $\langle N \rangle > 10^4$ ($R>5$~nm), but $S$ has been found to deviate from unity for smaller HNDs.~\cite{lewerenz1995successive,hessj1999measurement} To model the experimental ion yield, we assume the sticking coefficient to depend on the HND volume $V\propto N\propto R^3$ to account for the ability of HNDs to dissipate energy by evaporation of individual He atoms thereby increasing the capture probability. We model the size-depended sticking coefficient as
\begin{equation} \label{eq:stick}
    S(R) = 1 - e^{-(R/R_{\mathrm{s}})^3},
\end{equation}
where $R_s$ is a characteristic scale factor such that for $R=R_s$ the sticking coefficient is $1-1/e$. 
The probability to dope HNDs with $k=1$ or more dopant atoms or molecules is approximately given by Poissonian statistics
\begin{equation} \label{eq:poisson}
    P_{\mathrm{pickup}}(R) = 1 - P_{k=0} = 1 - e^{-n_l\sigma_{\mathrm{pickup}}(R)},
\end{equation}
where $n_l$ is the doping strength. Doping with alkali atoms has been found to deviate from Poissonian statistics~\cite{bunermann2011modeling}. However, for the sake of simplicity, we neglect this deviation here. At conditions where doping with one dopant atom is most likely, droplet shrinkage due to evaporation of He atoms after successful capture of a dopant atom can be neglected.~\cite{brink1990density,lehmann2004evaporative} 

The CT ionization probability of dopants has previously been studied and modelled.~\cite{ellis2007model,callicoatt1998capture,halberstadt1998resonant,lewis2005probing} Essentially the charge tends to move towards the dopant through a series of charge hopping events ($\sim 10$~hops). When the charge reaches the dopant in this series of hops, the dopant is ionized by CT. Assuming that O$_2$ sits in the center of the HND, the CT probability for an initial ionization at position ($r,\theta,\phi$) can be described as
\begin{equation} \label{eq:CT_O2_prob}
P_{\mathrm{CT}}^{\mathrm{(O_2)}}(r) = e^{-r/R_{\mathrm{CT}}},
\end{equation}
whereas the probability for CT ionization of Li sitting on the droplet surface is
\begin{equation} \label{eq:CT_Li_prob}
    P_{\mathrm{CT}}^{\mathrm{(Li)}}(r,\theta,R) = e^{-\Delta r/R_{\mathrm{CT}}},
\end{equation}
where $\Delta r = \sqrt{r^2+R^2-2rR\cos{\theta}}$ is the distance between He$^+$ and the Li atom; the coordinate system is oriented such that the Li atom is located at $\theta=0$. $R_{\mathrm{CT}}$ is the effective range of CT ionization in HNDs. In total, the relative dopant ion yield is then
\begin{equation} \label{eq:CT_O2}
    I_{\mathrm{CT}}^{\mathrm{(O_2)}}(R) = \frac{4\pi}{V}  P_{\mathrm{pickup}}(R) \int_0^R P_{\mathrm{CT}}^{\mathrm{(O_2)}}(r,R)r^2dr,
\end{equation}
\begin{equation} \label{eq:CT_Li}
    I_{\mathrm{CT}}^{\mathrm{(Li)}}(R) = \frac{2\pi}{V}  P_{\mathrm{pickup}}(R) \int_0^R \int_0^\pi P_{\mathrm{CT}}^{\mathrm{(Li)}}(r,\theta,R)r^2\sin\theta d\theta dr.
\end{equation}

\begin{table}[t]
\centering
\begin{tabular}{c|ccc}
 \multicolumn{1}{l}{Dopant ion} & \multicolumn{1}{l}{$n_l$ [$10^{-6}$~mbar$\times$m]} & \multicolumn{1}{l}{$R_S$ [nm]} & \multicolumn{1}{l}{$R_{\mathrm{CT}}$ [nm]} \\ \hline
O$_2^+$ & 0.8 & 3.8 & 2.0 \\
O$_2^+$ & 1.8 & 4.1 & 1.9 \\
Li$^+$ & 1.2 & 5.6 & 1.3
\end{tabular}
\caption{Fit results from CT ionization ($h\nu=26$~eV) of doped HNDs using the fit model given by Eq.~(\ref{eq:CT_O2}) and (\ref{eq:CT_Li}). $R_S$ is the effective droplet radius for capturing dopants and $R_{\mathrm{CT}}$ is the effective CT ionization range. The measured relative dopant ion yields and the fit curves are displayed in Fig.~\ref{fig:doped}~a).   }
\label{tab:CT}
\end{table}
Fig.~\ref{fig:doped}~a) shows the relative dopant ion yield as function of HND size, and Table~\ref{tab:CT} lists the resulting fit parameters. 
The determined fit parameter for the effective range of CT, $R_{\mathrm{CT}}$, matches the previous finding for a similar model, where the fit resulted in $R_{\mathrm{CT}}=0.8$-$3.5$~nm.~\cite{ellis2007model}
Note that the model does not contain any solvation effect of the dopant ion following CT ionization indicating that this effect is minor in the HND-size range where CT ionization is efficient. 
Most likely, solvation of ions is more prominent at larger HNDs; however, the probability of forming the dopant ion by CT ionization in the first place is small for large HNDs.
The negligible contribution of O$_2^+$ solvation for HND up to sizes $\langle R \rangle\sim 10^6$ ($R\sim20$~nm) is in contrast to the conclusion that He$_2^+$ is only ejected from the droplet if formed near the surface. 
O$_2^+$ ions formed in vibrationally excited states are more likely ejected by a prompt non-thermal process~\cite{yousif1987dissociative,smolarek2010ir}.
Since CT ionization is a weak channel for Li, the determined fit parameter is less reliable. 

Fig.~\ref{fig:doped}~b) shows the HND size-depended relative dopant ion yield from Penning ionization. The rising edges of the yields are similar to the case of CT ionization given by the probability to pick-up a dopant ($P_{\mathrm{pickup}}$). 
Unlike the CT ionization process, however, the dynamics of Penning ionization in HNDs has not been modelled to date. This is probably due to the fact that CT ionization is the typical ionization mechanism of dopants in mass spectrometers.~\cite{callicoatt1996charge,callicoatt1998capture,ruchti1998charge}
Additionally, while CT ionization is a purely electronic process happening on a time-scale $\sim$20~fs per charge hop~\cite{halberstadt1998resonant}, Penning ionization relies on nuclear dynamics in the superfluid medium happening on a time scale $>1$~ps.~\cite{mudrich2020ultrafast,laforge2021ultrafast,asmussen2022time} 
The initial excited state is delocalized in the HND resembling exciton states in heavier rare-gas clusters,~\cite{wormer1989fluorescence,stapelfeldt1989evolution} but localizes within $\sim100$~fs on a single atom or molecule (excimer).~\cite{closser2014simulations} Previous results indicated that interaction between dopants and the initial exciton-like state affects the localization favoring shorter interatomic distances between the dopant and the excited He atom.~\cite{asmussen2022time}
Thus, modelling of Penning ionization is more challenging. In this study, we make use of a simple model to describe the Penning ionization process. We assume that the excited He atom travels to the surface of the HND where Penning ionization may take place.~\cite{mudrich2020ultrafast} This means that Penning ionization is determined by two different processes in the case of doping with O$_2$ and Li residing in the droplet center and on the droplet surface, respectively. The probability of Penning ionization of O$_2$ depends only on the HND radius since this is the interparticle distance at which the decay happens assuming O$_2$ resides in the center of the droplet,
\begin{equation} \label{eq:P_O2_prob}
    P_{\mathrm{P}}^{\mathrm{(O_2)}}(R) = C_{\mathrm{P}}^{\mathrm{(O_2)}} e^{-R/R_\mathrm{P}}.
\end{equation}
Here, $C_{\mathrm{P}}^{\mathrm{(O_2)}}$ represents the product of the probability of the excited He atom to remain bound to the HND (instead of being ejected) and the probability of the dopant ion to be ejected (instead of being solvated in the droplets). When the HND size increases, the probability that the O$_2^+$ Penning ion is solvated also increases. Therefore, $R_{\mathrm{P}}$ is determined both by the effective range of Penning ionization and by solvation of the Penning ion in the droplet which also depends on the droplet size. Thus, the relative yield of ejected O$_2^+$ formed by Penning ionization is
\begin{equation} \label{eq:P_O2}
    I_{\mathrm{P}}^{\mathrm{(O_2)}}(R) = P_{\mathrm{pickup}}(R) P_{\mathrm{P}}^{\mathrm{(O_2)}}(R). 
\end{equation}
Table~\ref{tab:Penning} shows the results from fits of the HND size-depended Penning ionization $e$/O$_2^+$ coincidence yield.
From the fit, we determine that Penning ionization happens at an interatomic distance of $R_\mathrm{P} = 1$~nm. 
A previous study showed that Penning ionization is dominated by charge-exchange ICD occurring at an interatomic distance of $>5$~Å~\cite{ben2019charge}.  
In the case of Penning ionization of Li, both the excited He atom and the Li dopant reside at the HND surface which implies that the probability for Penning ionization is limited by the surface area covered by the two roaming species. In our model, we assume the Li atom to be frozen at the surface and the excited He atom roams about the droplet surface until it meets the Li atom. The probability for an excited He atom to ionize the Li is modelled as
\begin{equation} \label{eq:P_Li_prob}
    P_{\mathrm{P}}^{\mathrm{(Li)}}(R,\theta) = C_{\mathrm{P}}^{\mathrm{(Li)}} e^{-\Delta r/R_\mathrm{roam}},
\end{equation}
where $4\pi R_\mathrm{roam}^2$ is surface area explored by the roaming excited He atom and $\Delta r = \sqrt{2(1-\cos\theta)}R$ is the distance between He$^*$ and the Li atom. The relative Li$^+$ yield following Penning ionization is
\begin{equation} \label{eq:P_Li}
    I_{\mathrm{P}}^{\mathrm{(Li)}}(R) = \frac{1}{2} P_{\mathrm{pickup}}(R) \int_0^\pi P_{\mathrm{P}}^{\mathrm{(Li)}}(R,\theta) \sin\theta d\theta.
\end{equation}
This model reproduces qualitatively the finding that Penning ionization of Li leads to a significant yield of electron-ion coincidences even for large HNDs ($\langle N \rangle > 10^6$, $R>20$~nm), in contrast to Penning ionization of O$_2$ and CT ionization of both dopants (see Fig.~\ref{fig:doped}) where the ion yields vanishes. This model could be further improved by taking into account the excitation probability distribution in the HND; however, we refrain from extending our model in this study due to the limited experimental data available. Furthermore, we have recently demonstrated that electrons also suffer from losses due to their slowing down by elastic scattering, which eventually leads to electron solvation in bubbles and electron-ion recombination~\cite{asmussen2023electron}; this effect is not included in the present model.

Table~\ref{tab:Penning} summarizes the fit values for two different doping strengths. The large variation in the roaming parameter $R_\mathrm{roam}$ for the two fit curves indicates the limited validity of our simple model. The low value of the fit parameter $C_{\mathrm{P}}^{\mathrm{(Li)}} = 0.04$ found for both doping strengths shows that detection of Li Penning ions is relatively inefficient compared to other ionization mechanisms. This is most likely due to the efficient solvation of alkali cations in HNDs.~\cite{muller2009alkali,leal2014picosecond} 
\begin{table}[t]
\centering
\begin{tabular}{lllll}
\multicolumn{1}{l|}{Dopant ion} & $n_l$ [$10^{-6}$~mbar$\times$m] & $R_S$ [nm] & $C_{\mathrm{P}}^{\mathrm{(O_2)}}$ & $R_{\mathrm{P}}$ [nm] \\ \hline
\multicolumn{1}{l|}{O$_2^+$} & \multicolumn{1}{c}{1.8} & \multicolumn{1}{c}{4.0} & \multicolumn{1}{c}{0.36} & \multicolumn{1}{c}{1.1} \\
 &  &  &  &  \\
\multicolumn{1}{l|}{Dopant ion} & $n_l$ [$10^{-6}$~mbar$\times$m] & $R_S$ [nm] & $C_{\mathrm{P}}^{\mathrm{(Li)}}$ & $R_{\mathrm{roam}}$ [nm] \\ \hline
\multicolumn{1}{l|}{Li$^+$} & \multicolumn{1}{c}{0.9} & \multicolumn{1}{c}{7.4} & \multicolumn{1}{c}{0.04} & \multicolumn{1}{c}{102} \\
\multicolumn{1}{l|}{Li$^+$} & \multicolumn{1}{c}{1.2} & \multicolumn{1}{c}{4.2} & \multicolumn{1}{c}{0.04} & \multicolumn{1}{c}{66}
\end{tabular}
\caption{Fit results for Penning ionization ($h\nu=21.6$~eV) of doped HNDs using the models of Eqs.~(\ref{eq:P_O2}) and (\ref{eq:P_Li}). $R_{\mathrm{P}}$ and $R_{\mathrm{roam}}$ is the effective droplet size for which CT and Penning ionization is efficient, respectively. $C_{\mathrm{P}}$ is the product of the probability of the excited He atom to remain bound to the HND (instead of being ejected) and the probability of the dopant ion to be ejected (instead of being solvated in the droplets). The measured relative dopant ion yields and the fit curve are displayed in Fig.~\ref{fig:doped}~b).}
\label{tab:Penning}
\end{table}

The rising edge of the coincidence electron-dopant ion yield is relatively well reproduced by our simple model including the dopant capture probability. A discrepancy between the measured relative yields and the model is found for small HNDs where the yield tends to zero at larger HND radius compared to what the model predicts. This is most likely due to loss of HNDs under these conditions due to deflection of small HNDs after collisions with dopants and break-up of the HNDs by evaporation following capture of the dopants.  
In general, we find that the sticking coefficient ($S$) at a given HND size for O$_2$ ($R_S = 3.8$-$4.1$~nm) is larger than the one for Li ($R_S = 4.2$-$7.6$~nm). This means that 50\,\% of all dopant-droplet collisions lead to capture of the dopant ($S=0.5$) at a mean HND radius of $\sim$3~nm for O$_2$ and $\sim$4~nm for Li. A previous study of dopants bound in the droplet interior reported sticking coefficients ranging from 93\,\% to 52\,\% for various dopants with increasing mass at a droplet size of $\langle N \rangle=2650$.~\cite{lewerenz1995successive} However, for alkali atoms the sticking coefficient is expected to be lower compared to other dopants due to the weak binding of alkali atoms to the HND surface.~\cite{bunermann2011modeling,ancilotto1995sodium} For this reason, it is recommended to use larger HNDs for efficient doping of alkalis compared with other dopants. 
Penning ionization of alkalis on the surface of helium nanodroplets has previously been described as highly-efficient~\cite{buchta2013charge,ben2019charge} and is indeed more efficient compared to dopants residing in the HND bulk. However, here we show that the Li Penning ionization efficiency actually is low ($\lesssim 4$\,\%) when comparing with the total number of droplet excitations. At resonant excitation conditions where only Penning ionization of dopants leads to the emission of electrons and ions, though, the relatively low Penning ionization efficiency can readily be compensated by an increased photon flux.

\section{Conclusion}
In summary, we have systematically measured the yields of electrons, ions and electron-ion coincidences for pure and doped HNDs in a size range $\langle N \rangle = 10$-$10^9$ ($R=1$-$190$~nm). 
The main fragment ion ejected from ionized HNDs is He$_2^+$ followed by a series of He$_n^+$, $n=3,\,4,\dots$ clusters, whereas He$^+$ atomic ions are identified with the uncondensed, atomic component of the He jet. The latter rapidly drops as the HNDs grow larger from $\langle N \rangle<10^3$ up to $\langle N \rangle\approx 10^4$ ($R\approx5~$nm); In contrast, the yield of He$_2^+$ rises in this range of HND sizes. For larger HND sizes, He ions tend to remain bound to the HNDs where they form stable snowball structures inside the HNDs, leading to a reduction of the detected He ion yield. A simple model fitted to the measured He$_2^+$ yield reveals that only ionization near the droplet surface leads to the ejection and detection of He$_2^+$ ions. In contrast to ions, electrons are emitted from HNDs without losses up to larger sizes $\langle N \rangle\lesssim 10^6$ ($R \lesssim 25~$nm); Even larger HNDs tend to retain electrons, though, which impacts the detection of ions when detected in coincidence with electrons.

For HNDs doped with O$_2$ and Li, we have determined the optimal HND size where the probability of ionizing dopants and detecting the dopant ions is maximum, both in the regime of CT after ionization of the HNDs and for Penning ionization of the dopants after HND resonant excitation. The highest yields of O$_2^+$ and Li$^+$ from CT ionization are measured for a mean droplet size of $\langle N \rangle \approx 5\times10^4$. For Penning ionization of O$_2$ and Li, the highest ion yields are achieved for $\langle N \rangle \approx 10^3$ and $\langle N \rangle \approx 10^5$, respectively. Ionization of O$_2$ located in the HND interior is more efficient through CT ionization, whereas Li, residing on the HND surface, is most efficiently ionized by Penning ionization, as reported previously~\cite{buchta2013charge,an2011submersion}.
We rationalize the increased efficiency of Penning ionization of surface-bound dopants by a model where the excited He atom is expelled to the surface before Penning ionization. The CT ionization range is limited by the total hopping distance following ionization of the HND. Our model reveals a total hopping distance of $\sim 2$~nm, matching previous results~\cite{ellis2007model}. 
From simple fit models for the two ionization regimes, we infer the HND size-depended sticking coefficient for a dopant and the characteristic distance at which Penning/CT ionization takes place. We show that it requires a HND radius $R=4$~nm to ensure that 50\,\% of the droplet-Li collisions lead to successful capture whereas a radius $R=3$~nm is needed for 50\,\% capture probability for O$_2$.

These results can serve as guidelines to determine optimal experimental conditions for other experiments where efficient detection of ions and/or electrons formed by ionization of doped HNDs is important, such as mass spectrometry of dopant complexes aggregated in HNDs and IR/VIS/UV spectroscopy based on mass-spectrometric detection schemes.~\cite{messner2022photoabsorption}

\section*{Acknowledgement}
J.D.A. and M.M. acknowledge financial support by the Carlsberg Foundation. 
L.B.L. and M.M. acknowledge financial support by the Danish Council for Independent Research Fund (DFF) via Grant No. 1026-00299B.
S.R.K. thanks Dept. of Science and Technology, Govt. of India, for support through the DST-DAAD scheme and Science and Eng. Research Board. 
S.R.K. and K.S. acknowledge the support of the Scheme for Promotion of Academic Research Collaboration, Min. of Edu., Govt. of India, and the Institute of Excellence programme at IIT-Madras via the Quantum Center for Diamond and Emergent Materials. 
S.R.K. gratefully acknowledges support of the Max Planck Society's Partner group programme. 
M.M. and S.R.K. gratefully acknowledge funding from the SPARC Programme, MHRD, India. 
A.R.A. acknowledges with gratitude for the support from the Marie Skłodowska-Curie Postdoctoral Fellowship project Photochem-RS-RP (Grant Agreement No. 101068805) provided by the European Union’s Horizon 2020 Research and Innovation Programme.
The research leading to this result has been supported by the EU Framework Programme for Research and Innovation HORIZON 2020 and by the COST Action CA21101 ``Confined Molecular Systems: From a New Generation of Materials to the Stars (COSY)''.

\section*{Author declarations}
\subsection*{Conflict of Interest}
The authors have no conflicts to disclose.
\subsection*{Author Contributions}
J.D.A., L.B.L, K.S., A.R.A. and M.M. performed the experiments with support from B.B. and H.B.P.. S.K. aided remotely in the interpretation of the experimental results. J.D.A. and M.M. wrote the manuscript with input from all the co-authors.

\section*{References}
\nocite{*}
\bibliography{coinc_reference}

\begin{thebibliography}{95}%
\makeatletter
\providecommand \@ifxundefined [1]{%
 \@ifx{#1\undefined}
}%
\providecommand \@ifnum [1]{%
 \ifnum #1\expandafter \@firstoftwo
 \else \expandafter \@secondoftwo
 \fi
}%
\providecommand \@ifx [1]{%
 \ifx #1\expandafter \@firstoftwo
 \else \expandafter \@secondoftwo
 \fi
}%
\providecommand \natexlab [1]{#1}%
\providecommand \enquote  [1]{``#1''}%
\providecommand \bibnamefont  [1]{#1}%
\providecommand \bibfnamefont [1]{#1}%
\providecommand \citenamefont [1]{#1}%
\providecommand \href@noop [0]{\@secondoftwo}%
\providecommand \href [0]{\begingroup \@sanitize@url \@href}%
\providecommand \@href[1]{\@@startlink{#1}\@@href}%
\providecommand \@@href[1]{\endgroup#1\@@endlink}%
\providecommand \@sanitize@url [0]{\catcode `\\12\catcode `\$12\catcode
  `\&12\catcode `\#12\catcode `\^12\catcode `\_12\catcode `\%12\relax}%
\providecommand \@@startlink[1]{}%
\providecommand \@@endlink[0]{}%
\providecommand \url  [0]{\begingroup\@sanitize@url \@url }%
\providecommand \@url [1]{\endgroup\@href {#1}{\urlprefix }}%
\providecommand \urlprefix  [0]{URL }%
\providecommand \Eprint [0]{\href }%
\providecommand \doibase [0]{http://dx.doi.org/}%
\providecommand \selectlanguage [0]{\@gobble}%
\providecommand \bibinfo  [0]{\@secondoftwo}%
\providecommand \bibfield  [0]{\@secondoftwo}%
\providecommand \translation [1]{[#1]}%
\providecommand \BibitemOpen [0]{}%
\providecommand \bibitemStop [0]{}%
\providecommand \bibitemNoStop [0]{.\EOS\space}%
\providecommand \EOS [0]{\spacefactor3000\relax}%
\providecommand \BibitemShut  [1]{\csname bibitem#1\endcsname}%
\let\auto@bib@innerbib\@empty
\bibitem [{\citenamefont {Barranco}\ \emph {et~al.}(2006)\citenamefont
  {Barranco}, \citenamefont {Guardiola}, \citenamefont {Hern{\'a}ndez},
  \citenamefont {Mayol}, \citenamefont {Navarro},\ and\ \citenamefont
  {Pi}}]{barranco2006helium}%
  \BibitemOpen
  \bibfield  {author} {\bibinfo {author} {\bibfnamefont {M.}~\bibnamefont
  {Barranco}}, \bibinfo {author} {\bibfnamefont {R.}~\bibnamefont {Guardiola}},
  \bibinfo {author} {\bibfnamefont {S.}~\bibnamefont {Hern{\'a}ndez}}, \bibinfo
  {author} {\bibfnamefont {R.}~\bibnamefont {Mayol}}, \bibinfo {author}
  {\bibfnamefont {J.}~\bibnamefont {Navarro}}, \ and\ \bibinfo {author}
  {\bibfnamefont {M.}~\bibnamefont {Pi}},\ }\bibfield  {title} {\enquote
  {\bibinfo {title} {Helium nanodroplets: An overview},}\ }\href@noop {}
  {\bibfield  {journal} {\bibinfo  {journal} {J. Low Temp. Phys.}\ }\textbf
  {\bibinfo {volume} {142}},\ \bibinfo {pages} {1--81} (\bibinfo {year}
  {2006})}\BibitemShut {NoStop}%
\bibitem [{\citenamefont {Toennies}\ and\ \citenamefont
  {Vilesov}(2004)}]{toennies2004superfluid}%
  \BibitemOpen
  \bibfield  {author} {\bibinfo {author} {\bibfnamefont {J.~P.}\ \bibnamefont
  {Toennies}}\ and\ \bibinfo {author} {\bibfnamefont {A.~F.}\ \bibnamefont
  {Vilesov}},\ }\bibfield  {title} {\enquote {\bibinfo {title} {Superfluid
  helium droplets: A uniquely cold nanomatrix for molecules and molecular
  complexes},}\ }\href@noop {} {\bibfield  {journal} {\bibinfo  {journal}
  {Angewandte Chemie International Edition}\ }\textbf {\bibinfo {volume}
  {43}},\ \bibinfo {pages} {2622--2648} (\bibinfo {year} {2004})}\BibitemShut
  {NoStop}%
\bibitem [{\citenamefont {Mudrich}\ and\ \citenamefont
  {Stienkemeier}(2014)}]{mudrich2014photoionisaton}%
  \BibitemOpen
  \bibfield  {author} {\bibinfo {author} {\bibfnamefont {M.}~\bibnamefont
  {Mudrich}}\ and\ \bibinfo {author} {\bibfnamefont {F.}~\bibnamefont
  {Stienkemeier}},\ }\bibfield  {title} {\enquote {\bibinfo {title}
  {Photoionisaton of pure and doped helium nanodroplets},}\ }\href@noop {}
  {\bibfield  {journal} {\bibinfo  {journal} {International Reviews in Physical
  Chemistry}\ }\textbf {\bibinfo {volume} {33}},\ \bibinfo {pages} {301--339}
  (\bibinfo {year} {2014})}\BibitemShut {NoStop}%
\bibitem [{\citenamefont {Gonz{\'a}lez-Lezana}\ \emph
  {et~al.}(2020)\citenamefont {Gonz{\'a}lez-Lezana}, \citenamefont {Echt},
  \citenamefont {Gatchell}, \citenamefont {Bartolomei}, \citenamefont
  {Campos-Mart{\'\i}nez},\ and\ \citenamefont
  {Scheier}}]{gonzalez2020solvation}%
  \BibitemOpen
  \bibfield  {author} {\bibinfo {author} {\bibfnamefont {T.}~\bibnamefont
  {Gonz{\'a}lez-Lezana}}, \bibinfo {author} {\bibfnamefont {O.}~\bibnamefont
  {Echt}}, \bibinfo {author} {\bibfnamefont {M.}~\bibnamefont {Gatchell}},
  \bibinfo {author} {\bibfnamefont {M.}~\bibnamefont {Bartolomei}}, \bibinfo
  {author} {\bibfnamefont {J.}~\bibnamefont {Campos-Mart{\'\i}nez}}, \ and\
  \bibinfo {author} {\bibfnamefont {P.}~\bibnamefont {Scheier}},\ }\bibfield
  {title} {\enquote {\bibinfo {title} {Solvation of ions in helium},}\
  }\href@noop {} {\bibfield  {journal} {\bibinfo  {journal} {Int. Rev. Phys.
  Chem.}\ }\textbf {\bibinfo {volume} {39}},\ \bibinfo {pages} {465--516}
  (\bibinfo {year} {2020})}\BibitemShut {NoStop}%
\bibitem [{\citenamefont {Asmussen}\ \emph {et~al.}(2023)\citenamefont
  {Asmussen}, \citenamefont {Sishodia}, \citenamefont {Bastian}, \citenamefont
  {Abid}, \citenamefont {Ltaief}, \citenamefont {Pedersen}, \citenamefont {De},
  \citenamefont {Medina}, \citenamefont {Pal}, \citenamefont {Richter},
  \citenamefont {Fennel}, \citenamefont {Krishnan},\ and\ \citenamefont
  {Mudrich}}]{asmussen2023electron}%
  \BibitemOpen
  \bibfield  {author} {\bibinfo {author} {\bibfnamefont {J.~D.}\ \bibnamefont
  {Asmussen}}, \bibinfo {author} {\bibfnamefont {K.}~\bibnamefont {Sishodia}},
  \bibinfo {author} {\bibfnamefont {B.}~\bibnamefont {Bastian}}, \bibinfo
  {author} {\bibfnamefont {A.~R.}\ \bibnamefont {Abid}}, \bibinfo {author}
  {\bibfnamefont {L.~B.}\ \bibnamefont {Ltaief}}, \bibinfo {author}
  {\bibfnamefont {H.~B.}\ \bibnamefont {Pedersen}}, \bibinfo {author}
  {\bibfnamefont {S.}~\bibnamefont {De}}, \bibinfo {author} {\bibfnamefont
  {C.}~\bibnamefont {Medina}}, \bibinfo {author} {\bibfnamefont
  {N.}~\bibnamefont {Pal}}, \bibinfo {author} {\bibfnamefont {R.}~\bibnamefont
  {Richter}}, \bibinfo {author} {\bibfnamefont {T.}~\bibnamefont {Fennel}},
  \bibinfo {author} {\bibfnamefont {S.}~\bibnamefont {Krishnan}}, \ and\
  \bibinfo {author} {\bibfnamefont {M.}~\bibnamefont {Mudrich}},\ }\href@noop
  {} {\enquote {\bibinfo {title} {Electron energy loss and angular asymmetry
  induced by elastic scattering in helium droplets},}\ } (\bibinfo {year}
  {2023}),\ \Eprint {http://arxiv.org/abs/2305.05269} {arXiv:2305.05269
  [physics.atm-clus]} \BibitemShut {NoStop}%
\bibitem [{\citenamefont {Mudrich}\ \emph {et~al.}(2004)\citenamefont
  {Mudrich}, \citenamefont {B{\"u}nermann}, \citenamefont {Stienkemeier},
  \citenamefont {Dulieu},\ and\ \citenamefont
  {Weidem{\"u}ller}}]{mudrich2004formation}%
  \BibitemOpen
  \bibfield  {author} {\bibinfo {author} {\bibfnamefont {M.}~\bibnamefont
  {Mudrich}}, \bibinfo {author} {\bibfnamefont {O.}~\bibnamefont
  {B{\"u}nermann}}, \bibinfo {author} {\bibfnamefont {F.}~\bibnamefont
  {Stienkemeier}}, \bibinfo {author} {\bibfnamefont {O.}~\bibnamefont
  {Dulieu}}, \ and\ \bibinfo {author} {\bibfnamefont {M.}~\bibnamefont
  {Weidem{\"u}ller}},\ }\bibfield  {title} {\enquote {\bibinfo {title}
  {Formation of cold bialkali dimers on helium nanodroplets},}\ }\href@noop {}
  {\bibfield  {journal} {\bibinfo  {journal} {Eur. Phys. J. D}\ }\textbf
  {\bibinfo {volume} {31}},\ \bibinfo {pages} {291--299} (\bibinfo {year}
  {2004})}\BibitemShut {NoStop}%
\bibitem [{\citenamefont {Giese}\ \emph {et~al.}(2011)\citenamefont {Giese},
  \citenamefont {Stienkemeier}, \citenamefont {Mudrich}, \citenamefont
  {Hauser},\ and\ \citenamefont {Ernst}}]{giese2011homo}%
  \BibitemOpen
  \bibfield  {author} {\bibinfo {author} {\bibfnamefont {C.}~\bibnamefont
  {Giese}}, \bibinfo {author} {\bibfnamefont {F.}~\bibnamefont {Stienkemeier}},
  \bibinfo {author} {\bibfnamefont {M.}~\bibnamefont {Mudrich}}, \bibinfo
  {author} {\bibfnamefont {A.~W.}\ \bibnamefont {Hauser}}, \ and\ \bibinfo
  {author} {\bibfnamefont {W.~E.}\ \bibnamefont {Ernst}},\ }\bibfield  {title}
  {\enquote {\bibinfo {title} {Homo-and heteronuclear alkali metal trimers
  formed on helium nanodroplets. {P}art {II}. {F}emtosecond spectroscopy and
  spectra assignments},}\ }\href@noop {} {\bibfield  {journal} {\bibinfo
  {journal} {Phys. Chem. Chem. Phys.}\ }\textbf {\bibinfo {volume} {13}},\
  \bibinfo {pages} {18769--18780} (\bibinfo {year} {2011})}\BibitemShut
  {NoStop}%
\bibitem [{\citenamefont {Loginov}, \citenamefont {Braun},\ and\ \citenamefont
  {Drabbels}(2008)}]{loginov2008new}%
  \BibitemOpen
  \bibfield  {author} {\bibinfo {author} {\bibfnamefont {E.}~\bibnamefont
  {Loginov}}, \bibinfo {author} {\bibfnamefont {A.}~\bibnamefont {Braun}}, \
  and\ \bibinfo {author} {\bibfnamefont {M.}~\bibnamefont {Drabbels}},\
  }\bibfield  {title} {\enquote {\bibinfo {title} {A new sensitive detection
  scheme for helium nanodroplet isolation spectroscopy: {A}pplication to
  benzene},}\ }\href@noop {} {\bibfield  {journal} {\bibinfo  {journal} {Phys.
  Chem. Chem. Phys.}\ }\textbf {\bibinfo {volume} {10}},\ \bibinfo {pages}
  {6107--6114} (\bibinfo {year} {2008})}\BibitemShut {NoStop}%
\bibitem [{\citenamefont {Smolarek}\ \emph {et~al.}(2010)\citenamefont
  {Smolarek}, \citenamefont {Brauer}, \citenamefont {Buma},\ and\ \citenamefont
  {Drabbels}}]{smolarek2010ir}%
  \BibitemOpen
  \bibfield  {author} {\bibinfo {author} {\bibfnamefont {S.}~\bibnamefont
  {Smolarek}}, \bibinfo {author} {\bibfnamefont {N.~B.}\ \bibnamefont
  {Brauer}}, \bibinfo {author} {\bibfnamefont {W.~J.}\ \bibnamefont {Buma}}, \
  and\ \bibinfo {author} {\bibfnamefont {M.}~\bibnamefont {Drabbels}},\
  }\bibfield  {title} {\enquote {\bibinfo {title} {{IR} spectroscopy of
  molecular ions by nonthermal ion ejection from helium nanodroplets},}\
  }\href@noop {} {\ \textbf {\bibinfo {volume} {132}},\ \bibinfo {pages}
  {14086--14091} (\bibinfo {year} {2010})}\BibitemShut {NoStop}%
\bibitem [{\citenamefont {Tiggesb{\"a}umker}\ and\ \citenamefont
  {Stienkemeier}(2007)}]{tiggesbaumker2007formation}%
  \BibitemOpen
  \bibfield  {author} {\bibinfo {author} {\bibfnamefont {J.}~\bibnamefont
  {Tiggesb{\"a}umker}}\ and\ \bibinfo {author} {\bibfnamefont {F.}~\bibnamefont
  {Stienkemeier}},\ }\bibfield  {title} {\enquote {\bibinfo {title} {Formation
  and properties of metal clusters isolated in helium droplets},}\ }\href@noop
  {} {\bibfield  {journal} {\bibinfo  {journal} {Phys. Chem. Chem. Phys.}\
  }\textbf {\bibinfo {volume} {9}},\ \bibinfo {pages} {4748--4770} (\bibinfo
  {year} {2007})}\BibitemShut {NoStop}%
\bibitem [{\citenamefont {Theisen}, \citenamefont {Lackner},\ and\
  \citenamefont {Ernst}(2011)}]{theisen2011rb}%
  \BibitemOpen
  \bibfield  {author} {\bibinfo {author} {\bibfnamefont {M.}~\bibnamefont
  {Theisen}}, \bibinfo {author} {\bibfnamefont {F.}~\bibnamefont {Lackner}}, \
  and\ \bibinfo {author} {\bibfnamefont {W.~E.}\ \bibnamefont {Ernst}},\
  }\bibfield  {title} {\enquote {\bibinfo {title} {Rb and {C}s oligomers in
  different spin configurations on helium nanodroplets},}\ }\href@noop {}
  {\bibfield  {journal} {\bibinfo  {journal} {J. Phys. Chem. A}\ }\textbf
  {\bibinfo {volume} {115}},\ \bibinfo {pages} {7005--7009} (\bibinfo {year}
  {2011})}\BibitemShut {NoStop}%
\bibitem [{\citenamefont {Kazak}\ \emph {et~al.}(2019)\citenamefont {Kazak},
  \citenamefont {G{\"o}de}, \citenamefont {Meiwes-Broer},\ and\ \citenamefont
  {Tiggesb{\"a}umker}}]{kazak2019photoelectron}%
  \BibitemOpen
  \bibfield  {author} {\bibinfo {author} {\bibfnamefont {L.}~\bibnamefont
  {Kazak}}, \bibinfo {author} {\bibfnamefont {S.}~\bibnamefont {G{\"o}de}},
  \bibinfo {author} {\bibfnamefont {K.-H.}\ \bibnamefont {Meiwes-Broer}}, \
  and\ \bibinfo {author} {\bibfnamefont {J.}~\bibnamefont
  {Tiggesb{\"a}umker}},\ }\bibfield  {title} {\enquote {\bibinfo {title}
  {Photoelectron spectroscopy on magnesium ensembles in helium nanodroplets},}\
  }\href@noop {} {\ \textbf {\bibinfo {volume} {123}},\ \bibinfo {pages}
  {5951--5956} (\bibinfo {year} {2019})}\BibitemShut {NoStop}%
\bibitem [{\citenamefont {Messner}\ \emph {et~al.}(2018)\citenamefont
  {Messner}, \citenamefont {Schiffmann}, \citenamefont {Pototschnig},
  \citenamefont {Lasserus}, \citenamefont {Schnedlitz}, \citenamefont
  {Lackner},\ and\ \citenamefont {Ernst}}]{messner2018spectroscopy}%
  \BibitemOpen
  \bibfield  {author} {\bibinfo {author} {\bibfnamefont {R.}~\bibnamefont
  {Messner}}, \bibinfo {author} {\bibfnamefont {A.}~\bibnamefont {Schiffmann}},
  \bibinfo {author} {\bibfnamefont {J.~V.}\ \bibnamefont {Pototschnig}},
  \bibinfo {author} {\bibfnamefont {M.}~\bibnamefont {Lasserus}}, \bibinfo
  {author} {\bibfnamefont {M.}~\bibnamefont {Schnedlitz}}, \bibinfo {author}
  {\bibfnamefont {F.}~\bibnamefont {Lackner}}, \ and\ \bibinfo {author}
  {\bibfnamefont {W.~E.}\ \bibnamefont {Ernst}},\ }\bibfield  {title} {\enquote
  {\bibinfo {title} {Spectroscopy of gold atoms and gold oligomers in helium
  nanodroplets},}\ }\href@noop {} {\bibfield  {journal} {\bibinfo  {journal}
  {J. Chem. Phys.}\ }\textbf {\bibinfo {volume} {149}},\ \bibinfo {pages}
  {024305} (\bibinfo {year} {2018})}\BibitemShut {NoStop}%
\bibitem [{\citenamefont {Braun}\ and\ \citenamefont
  {Drabbels}(2004)}]{braun2004imaging}%
  \BibitemOpen
  \bibfield  {author} {\bibinfo {author} {\bibfnamefont {A.}~\bibnamefont
  {Braun}}\ and\ \bibinfo {author} {\bibfnamefont {M.}~\bibnamefont
  {Drabbels}},\ }\bibfield  {title} {\enquote {\bibinfo {title} {Imaging the
  translational dynamics of {CF}3 in liquid helium droplets},}\ }\href@noop {}
  {\bibfield  {journal} {\bibinfo  {journal} {Phys. Rev. Lett.}\ }\textbf
  {\bibinfo {volume} {93}},\ \bibinfo {pages} {253401} (\bibinfo {year}
  {2004})}\BibitemShut {NoStop}%
\bibitem [{\citenamefont {Giese}\ \emph {et~al.}(2012)\citenamefont {Giese},
  \citenamefont {Mullins}, \citenamefont {Gr{\"u}ner}, \citenamefont
  {Weidem{\"u}ller}, \citenamefont {Stienkemeier},\ and\ \citenamefont
  {Mudrich}}]{giese2012formation}%
  \BibitemOpen
  \bibfield  {author} {\bibinfo {author} {\bibfnamefont {C.}~\bibnamefont
  {Giese}}, \bibinfo {author} {\bibfnamefont {T.}~\bibnamefont {Mullins}},
  \bibinfo {author} {\bibfnamefont {B.}~\bibnamefont {Gr{\"u}ner}}, \bibinfo
  {author} {\bibfnamefont {M.}~\bibnamefont {Weidem{\"u}ller}}, \bibinfo
  {author} {\bibfnamefont {F.}~\bibnamefont {Stienkemeier}}, \ and\ \bibinfo
  {author} {\bibfnamefont {M.}~\bibnamefont {Mudrich}},\ }\bibfield  {title}
  {\enquote {\bibinfo {title} {Formation and relaxation of {R}b{H}e exciplexes
  on {H}e nanodroplets studied by femtosecond pump and picosecond probe
  spectroscopy},}\ }\href@noop {} {\bibfield  {journal} {\bibinfo  {journal}
  {J. Chem. Phys.}\ }\textbf {\bibinfo {volume} {137}},\ \bibinfo {pages}
  {024316} (\bibinfo {year} {2012})}\BibitemShut {NoStop}%
\bibitem [{\citenamefont {Kautsch}, \citenamefont {Koch},\ and\ \citenamefont
  {Ernst}(2015)}]{kautsch2015photoinduced}%
  \BibitemOpen
  \bibfield  {author} {\bibinfo {author} {\bibfnamefont {A.}~\bibnamefont
  {Kautsch}}, \bibinfo {author} {\bibfnamefont {M.}~\bibnamefont {Koch}}, \
  and\ \bibinfo {author} {\bibfnamefont {W.~E.}\ \bibnamefont {Ernst}},\
  }\bibfield  {title} {\enquote {\bibinfo {title} {Photoinduced molecular
  dissociation and photoinduced recombination mediated by superfluid helium
  nanodroplets},}\ }\href@noop {} {\bibfield  {journal} {\bibinfo  {journal}
  {Phys. Chem. Chem. Phys.}\ }\textbf {\bibinfo {volume} {17}},\ \bibinfo
  {pages} {12310--12316} (\bibinfo {year} {2015})}\BibitemShut {NoStop}%
\bibitem [{\citenamefont {von Vangerow}\ \emph {et~al.}(2017)\citenamefont {von
  Vangerow}, \citenamefont {Coppens}, \citenamefont {Leal}, \citenamefont {Pi},
  \citenamefont {Barranco}, \citenamefont {Halberstadt}, \citenamefont
  {Stienkemeier},\ and\ \citenamefont {Mudrich}}]{von2017imaging}%
  \BibitemOpen
  \bibfield  {author} {\bibinfo {author} {\bibfnamefont {J.}~\bibnamefont {von
  Vangerow}}, \bibinfo {author} {\bibfnamefont {F.}~\bibnamefont {Coppens}},
  \bibinfo {author} {\bibfnamefont {A.}~\bibnamefont {Leal}}, \bibinfo {author}
  {\bibfnamefont {M.}~\bibnamefont {Pi}}, \bibinfo {author} {\bibfnamefont
  {M.}~\bibnamefont {Barranco}}, \bibinfo {author} {\bibfnamefont
  {N.}~\bibnamefont {Halberstadt}}, \bibinfo {author} {\bibfnamefont
  {F.}~\bibnamefont {Stienkemeier}}, \ and\ \bibinfo {author} {\bibfnamefont
  {M.}~\bibnamefont {Mudrich}},\ }\bibfield  {title} {\enquote {\bibinfo
  {title} {Imaging excited-state dynamics of doped {H}e nanodroplets in
  real-time},}\ }\href@noop {} {\bibfield  {journal} {\bibinfo  {journal} {J.
  Phys. Chem. Lett.}\ }\textbf {\bibinfo {volume} {8}},\ \bibinfo {pages}
  {307--312} (\bibinfo {year} {2017})}\BibitemShut {NoStop}%
\bibitem [{\citenamefont {Thaler}\ \emph {et~al.}(2020)\citenamefont {Thaler},
  \citenamefont {Heim}, \citenamefont {Treiber},\ and\ \citenamefont
  {Koch}}]{thaler2020ultrafast}%
  \BibitemOpen
  \bibfield  {author} {\bibinfo {author} {\bibfnamefont {B.}~\bibnamefont
  {Thaler}}, \bibinfo {author} {\bibfnamefont {P.}~\bibnamefont {Heim}},
  \bibinfo {author} {\bibfnamefont {L.}~\bibnamefont {Treiber}}, \ and\
  \bibinfo {author} {\bibfnamefont {M.}~\bibnamefont {Koch}},\ }\bibfield
  {title} {\enquote {\bibinfo {title} {Ultrafast photoinduced dynamics of
  single atoms solvated inside helium nanodroplets},}\ }\href@noop {}
  {\bibfield  {journal} {\bibinfo  {journal} {J. Chem. Phys.}\ }\textbf
  {\bibinfo {volume} {152}},\ \bibinfo {pages} {014307} (\bibinfo {year}
  {2020})}\BibitemShut {NoStop}%
\bibitem [{\citenamefont {Stadlhofer}, \citenamefont {Thaler},\ and\
  \citenamefont {Koch}(2022)}]{stadlhofer2022dimer}%
  \BibitemOpen
  \bibfield  {author} {\bibinfo {author} {\bibfnamefont {M.}~\bibnamefont
  {Stadlhofer}}, \bibinfo {author} {\bibfnamefont {B.}~\bibnamefont {Thaler}},
  \ and\ \bibinfo {author} {\bibfnamefont {M.}~\bibnamefont {Koch}},\
  }\bibfield  {title} {\enquote {\bibinfo {title} {Dimer photofragmentation and
  cation ejection dynamics in helium nanodroplets},}\ }\href@noop {} {\bibfield
   {journal} {\bibinfo  {journal} {Phys. Chem. Chem. Phys.}\ }\textbf {\bibinfo
  {volume} {24}},\ \bibinfo {pages} {24727--24733} (\bibinfo {year}
  {2022})}\BibitemShut {NoStop}%
\bibitem [{\citenamefont {Asmussen}\ \emph {et~al.}(2022)\citenamefont
  {Asmussen}, \citenamefont {Michiels}, \citenamefont {Bangert}, \citenamefont
  {Sisourat}, \citenamefont {Binz}, \citenamefont {Bruder}, \citenamefont
  {Danailov}, \citenamefont {Di~Fraia}, \citenamefont {Feifel}, \citenamefont
  {Giannessi} \emph {et~al.}}]{asmussen2022time}%
  \BibitemOpen
  \bibfield  {author} {\bibinfo {author} {\bibfnamefont {J.~D.}\ \bibnamefont
  {Asmussen}}, \bibinfo {author} {\bibfnamefont {R.}~\bibnamefont {Michiels}},
  \bibinfo {author} {\bibfnamefont {U.}~\bibnamefont {Bangert}}, \bibinfo
  {author} {\bibfnamefont {N.}~\bibnamefont {Sisourat}}, \bibinfo {author}
  {\bibfnamefont {M.}~\bibnamefont {Binz}}, \bibinfo {author} {\bibfnamefont
  {L.}~\bibnamefont {Bruder}}, \bibinfo {author} {\bibfnamefont
  {M.}~\bibnamefont {Danailov}}, \bibinfo {author} {\bibfnamefont
  {M.}~\bibnamefont {Di~Fraia}}, \bibinfo {author} {\bibfnamefont
  {R.}~\bibnamefont {Feifel}}, \bibinfo {author} {\bibfnamefont
  {L.}~\bibnamefont {Giannessi}},  \emph {et~al.},\ }\bibfield  {title}
  {\enquote {\bibinfo {title} {Time-resolved ultrafast interatomic coulombic
  decay in superexcited sodium-doped helium nanodroplets},}\ }\href@noop {}
  {\bibfield  {journal} {\bibinfo  {journal} {J. Phys. Chem. Lett.}\ }\textbf
  {\bibinfo {volume} {13}},\ \bibinfo {pages} {4470--4478} (\bibinfo {year}
  {2022})}\BibitemShut {NoStop}%
\bibitem [{\citenamefont {Haberfehlner}\ \emph {et~al.}(2015)\citenamefont
  {Haberfehlner}, \citenamefont {Thaler}, \citenamefont {Knez}, \citenamefont
  {Volk}, \citenamefont {Hofer}, \citenamefont {Ernst},\ and\ \citenamefont
  {Kothleitner}}]{haberfehlner2015formation}%
  \BibitemOpen
  \bibfield  {author} {\bibinfo {author} {\bibfnamefont {G.}~\bibnamefont
  {Haberfehlner}}, \bibinfo {author} {\bibfnamefont {P.}~\bibnamefont
  {Thaler}}, \bibinfo {author} {\bibfnamefont {D.}~\bibnamefont {Knez}},
  \bibinfo {author} {\bibfnamefont {A.}~\bibnamefont {Volk}}, \bibinfo {author}
  {\bibfnamefont {F.}~\bibnamefont {Hofer}}, \bibinfo {author} {\bibfnamefont
  {W.~E.}\ \bibnamefont {Ernst}}, \ and\ \bibinfo {author} {\bibfnamefont
  {G.}~\bibnamefont {Kothleitner}},\ }\bibfield  {title} {\enquote {\bibinfo
  {title} {Formation of bimetallic clusters in superfluid helium nanodroplets
  analysed by atomic resolution electron tomography},}\ }\href@noop {}
  {\bibfield  {journal} {\bibinfo  {journal} {Nature Communications}\ }\textbf
  {\bibinfo {volume} {6}},\ \bibinfo {pages} {8779} (\bibinfo {year}
  {2015})}\BibitemShut {NoStop}%
\bibitem [{\citenamefont {Messner}, \citenamefont {Ernst},\ and\ \citenamefont
  {Lackner}(2020)}]{messner2020shell}%
  \BibitemOpen
  \bibfield  {author} {\bibinfo {author} {\bibfnamefont {R.}~\bibnamefont
  {Messner}}, \bibinfo {author} {\bibfnamefont {W.~E.}\ \bibnamefont {Ernst}},
  \ and\ \bibinfo {author} {\bibfnamefont {F.}~\bibnamefont {Lackner}},\
  }\bibfield  {title} {\enquote {\bibinfo {title} {Shell-isolated {A}u
  nanoparticles functionalized with rhodamine {B} fluorophores in helium
  nanodroplets},}\ }\href@noop {} {\bibfield  {journal} {\bibinfo  {journal}
  {J. Phys. Chem. Lett.}\ }\textbf {\bibinfo {volume} {12}},\ \bibinfo {pages}
  {145--150} (\bibinfo {year} {2020})}\BibitemShut {NoStop}%
\bibitem [{\citenamefont {Ali{\'c}}\ \emph {et~al.}(2023)\citenamefont
  {Ali{\'c}}, \citenamefont {Messner}, \citenamefont {Ale{\v{s}}kovi{\'c}},
  \citenamefont {K{\"u}stner}, \citenamefont {Rub{\v{c}}i{\'c}}, \citenamefont
  {Lackner}, \citenamefont {Ernst},\ and\ \citenamefont
  {{\v{S}}ekutor}}]{alic2023diamondoid}%
  \BibitemOpen
  \bibfield  {author} {\bibinfo {author} {\bibfnamefont {J.}~\bibnamefont
  {Ali{\'c}}}, \bibinfo {author} {\bibfnamefont {R.}~\bibnamefont {Messner}},
  \bibinfo {author} {\bibfnamefont {M.}~\bibnamefont {Ale{\v{s}}kovi{\'c}}},
  \bibinfo {author} {\bibfnamefont {F.}~\bibnamefont {K{\"u}stner}}, \bibinfo
  {author} {\bibfnamefont {M.}~\bibnamefont {Rub{\v{c}}i{\'c}}}, \bibinfo
  {author} {\bibfnamefont {F.}~\bibnamefont {Lackner}}, \bibinfo {author}
  {\bibfnamefont {W.~E.}\ \bibnamefont {Ernst}}, \ and\ \bibinfo {author}
  {\bibfnamefont {M.}~\bibnamefont {{\v{S}}ekutor}},\ }\bibfield  {title}
  {\enquote {\bibinfo {title} {Diamondoid ether clusters in helium
  nanodroplets},}\ }\href@noop {} {\bibfield  {journal} {\bibinfo  {journal}
  {Phys. Chem. Chem. Phys.}\ } (\bibinfo {year} {2023})}\BibitemShut {NoStop}%
\bibitem [{\citenamefont {Schiffmann}\ \emph {et~al.}(2020)\citenamefont
  {Schiffmann}, \citenamefont {Jauk}, \citenamefont {Knez}, \citenamefont
  {Fitzek}, \citenamefont {Hofer}, \citenamefont {Lackner},\ and\ \citenamefont
  {Ernst}}]{schiffmann2020helium}%
  \BibitemOpen
  \bibfield  {author} {\bibinfo {author} {\bibfnamefont {A.}~\bibnamefont
  {Schiffmann}}, \bibinfo {author} {\bibfnamefont {T.}~\bibnamefont {Jauk}},
  \bibinfo {author} {\bibfnamefont {D.}~\bibnamefont {Knez}}, \bibinfo {author}
  {\bibfnamefont {H.}~\bibnamefont {Fitzek}}, \bibinfo {author} {\bibfnamefont
  {F.}~\bibnamefont {Hofer}}, \bibinfo {author} {\bibfnamefont
  {F.}~\bibnamefont {Lackner}}, \ and\ \bibinfo {author} {\bibfnamefont
  {W.~E.}\ \bibnamefont {Ernst}},\ }\bibfield  {title} {\enquote {\bibinfo
  {title} {Helium droplet assisted synthesis of plasmonic {A}g@{Z}no core@shell
  nanoparticles},}\ }\href@noop {} {\bibfield  {journal} {\bibinfo  {journal}
  {Nano Research}\ }\textbf {\bibinfo {volume} {13}},\ \bibinfo {pages}
  {2979--2986} (\bibinfo {year} {2020})}\BibitemShut {NoStop}%
\bibitem [{\citenamefont {Martini}\ \emph {et~al.}(2021)\citenamefont
  {Martini}, \citenamefont {Albertini}, \citenamefont {Laimer}, \citenamefont
  {Meyer}, \citenamefont {Gatchell}, \citenamefont {Echt}, \citenamefont
  {Zappa},\ and\ \citenamefont {Scheier}}]{martini2021splashing}%
  \BibitemOpen
  \bibfield  {author} {\bibinfo {author} {\bibfnamefont {P.}~\bibnamefont
  {Martini}}, \bibinfo {author} {\bibfnamefont {S.}~\bibnamefont {Albertini}},
  \bibinfo {author} {\bibfnamefont {F.}~\bibnamefont {Laimer}}, \bibinfo
  {author} {\bibfnamefont {M.}~\bibnamefont {Meyer}}, \bibinfo {author}
  {\bibfnamefont {M.}~\bibnamefont {Gatchell}}, \bibinfo {author}
  {\bibfnamefont {O.}~\bibnamefont {Echt}}, \bibinfo {author} {\bibfnamefont
  {F.}~\bibnamefont {Zappa}}, \ and\ \bibinfo {author} {\bibfnamefont
  {P.}~\bibnamefont {Scheier}},\ }\bibfield  {title} {\enquote {\bibinfo
  {title} {Splashing of large helium nanodroplets upon surface collisions},}\
  }\href@noop {} {\bibfield  {journal} {\bibinfo  {journal} {Phys. Rev. Lett.}\
  }\textbf {\bibinfo {volume} {127}},\ \bibinfo {pages} {263401} (\bibinfo
  {year} {2021})}\BibitemShut {NoStop}%
\bibitem [{\citenamefont {Gomez}\ \emph {et~al.}(2014)\citenamefont {Gomez},
  \citenamefont {Ferguson}, \citenamefont {Cryan}, \citenamefont {Bacellar},
  \citenamefont {Tanyag}, \citenamefont {Jones}, \citenamefont {Schorb},
  \citenamefont {Anielski}, \citenamefont {Belkacem}, \citenamefont {Bernando}
  \emph {et~al.}}]{gomez2014shapes}%
  \BibitemOpen
  \bibfield  {author} {\bibinfo {author} {\bibfnamefont {L.~F.}\ \bibnamefont
  {Gomez}}, \bibinfo {author} {\bibfnamefont {K.~R.}\ \bibnamefont {Ferguson}},
  \bibinfo {author} {\bibfnamefont {J.~P.}\ \bibnamefont {Cryan}}, \bibinfo
  {author} {\bibfnamefont {C.}~\bibnamefont {Bacellar}}, \bibinfo {author}
  {\bibfnamefont {R.~M.~P.}\ \bibnamefont {Tanyag}}, \bibinfo {author}
  {\bibfnamefont {C.}~\bibnamefont {Jones}}, \bibinfo {author} {\bibfnamefont
  {S.}~\bibnamefont {Schorb}}, \bibinfo {author} {\bibfnamefont
  {D.}~\bibnamefont {Anielski}}, \bibinfo {author} {\bibfnamefont
  {A.}~\bibnamefont {Belkacem}}, \bibinfo {author} {\bibfnamefont
  {C.}~\bibnamefont {Bernando}},  \emph {et~al.},\ }\bibfield  {title}
  {\enquote {\bibinfo {title} {Shapes and vorticities of superfluid helium
  nanodroplets},}\ }\href@noop {} {\bibfield  {journal} {\bibinfo  {journal}
  {Science}\ }\textbf {\bibinfo {volume} {345}},\ \bibinfo {pages} {906--909}
  (\bibinfo {year} {2014})}\BibitemShut {NoStop}%
\bibitem [{\citenamefont {Langbehn}\ \emph {et~al.}(2022)\citenamefont
  {Langbehn}, \citenamefont {Ovcharenko}, \citenamefont {Clark}, \citenamefont
  {Coreno}, \citenamefont {Cucini}, \citenamefont {Demidovich}, \citenamefont
  {Drabbels}, \citenamefont {Finetti}, \citenamefont {Di~Fraia}, \citenamefont
  {Giannessi} \emph {et~al.}}]{langbehn2022diffraction}%
  \BibitemOpen
  \bibfield  {author} {\bibinfo {author} {\bibfnamefont {B.}~\bibnamefont
  {Langbehn}}, \bibinfo {author} {\bibfnamefont {Y.}~\bibnamefont
  {Ovcharenko}}, \bibinfo {author} {\bibfnamefont {A.}~\bibnamefont {Clark}},
  \bibinfo {author} {\bibfnamefont {M.}~\bibnamefont {Coreno}}, \bibinfo
  {author} {\bibfnamefont {R.}~\bibnamefont {Cucini}}, \bibinfo {author}
  {\bibfnamefont {A.}~\bibnamefont {Demidovich}}, \bibinfo {author}
  {\bibfnamefont {M.}~\bibnamefont {Drabbels}}, \bibinfo {author}
  {\bibfnamefont {P.}~\bibnamefont {Finetti}}, \bibinfo {author} {\bibfnamefont
  {M.}~\bibnamefont {Di~Fraia}}, \bibinfo {author} {\bibfnamefont
  {L.}~\bibnamefont {Giannessi}},  \emph {et~al.},\ }\bibfield  {title}
  {\enquote {\bibinfo {title} {Diffraction imaging of light induced dynamics in
  xenon-doped helium nanodroplets},}\ }\href@noop {} {\bibfield  {journal}
  {\bibinfo  {journal} {New Journal of Physics}\ }\textbf {\bibinfo {volume}
  {24}},\ \bibinfo {pages} {113043} (\bibinfo {year} {2022})}\BibitemShut
  {NoStop}%
\bibitem [{\citenamefont {Thaler}\ \emph {et~al.}(2015)\citenamefont {Thaler},
  \citenamefont {Volk}, \citenamefont {Knez}, \citenamefont {Lackner},
  \citenamefont {Haberfehlner}, \citenamefont {Steurer}, \citenamefont
  {Schnedlitz},\ and\ \citenamefont {Ernst}}]{thaler2015synthesis}%
  \BibitemOpen
  \bibfield  {author} {\bibinfo {author} {\bibfnamefont {P.}~\bibnamefont
  {Thaler}}, \bibinfo {author} {\bibfnamefont {A.}~\bibnamefont {Volk}},
  \bibinfo {author} {\bibfnamefont {D.}~\bibnamefont {Knez}}, \bibinfo {author}
  {\bibfnamefont {F.}~\bibnamefont {Lackner}}, \bibinfo {author} {\bibfnamefont
  {G.}~\bibnamefont {Haberfehlner}}, \bibinfo {author} {\bibfnamefont
  {J.}~\bibnamefont {Steurer}}, \bibinfo {author} {\bibfnamefont
  {M.}~\bibnamefont {Schnedlitz}}, \ and\ \bibinfo {author} {\bibfnamefont
  {W.~E.}\ \bibnamefont {Ernst}},\ }\bibfield  {title} {\enquote {\bibinfo
  {title} {Synthesis of nanoparticles in helium droplets — a characterization
  comparing mass-spectra and electron microscopy data},}\ }\href@noop {}
  {\bibfield  {journal} {\bibinfo  {journal} {J. Chem. Phys.}\ }\textbf
  {\bibinfo {volume} {143}},\ \bibinfo {pages} {134201} (\bibinfo {year}
  {2015})}\BibitemShut {NoStop}%
\bibitem [{\citenamefont {Mauracher}\ \emph {et~al.}(2018)\citenamefont
  {Mauracher}, \citenamefont {Echt}, \citenamefont {Ellis}, \citenamefont
  {Yang}, \citenamefont {Bohme}, \citenamefont {Postler}, \citenamefont
  {Kaiser}, \citenamefont {Denifl},\ and\ \citenamefont
  {Scheier}}]{mauracher2018cold}%
  \BibitemOpen
  \bibfield  {author} {\bibinfo {author} {\bibfnamefont {A.}~\bibnamefont
  {Mauracher}}, \bibinfo {author} {\bibfnamefont {O.}~\bibnamefont {Echt}},
  \bibinfo {author} {\bibfnamefont {A.}~\bibnamefont {Ellis}}, \bibinfo
  {author} {\bibfnamefont {S.}~\bibnamefont {Yang}}, \bibinfo {author}
  {\bibfnamefont {D.}~\bibnamefont {Bohme}}, \bibinfo {author} {\bibfnamefont
  {J.}~\bibnamefont {Postler}}, \bibinfo {author} {\bibfnamefont
  {A.}~\bibnamefont {Kaiser}}, \bibinfo {author} {\bibfnamefont
  {S.}~\bibnamefont {Denifl}}, \ and\ \bibinfo {author} {\bibfnamefont
  {P.}~\bibnamefont {Scheier}},\ }\bibfield  {title} {\enquote {\bibinfo
  {title} {Cold physics and chemistry: {C}ollisions, ionization and reactions
  inside helium nanodroplets close to zero {K}},}\ }\href@noop {} {\bibfield
  {journal} {\bibinfo  {journal} {Physics Reports}\ }\textbf {\bibinfo {volume}
  {751}},\ \bibinfo {pages} {1--90} (\bibinfo {year} {2018})}\BibitemShut
  {NoStop}%
\bibitem [{\citenamefont {Atkins}(1959)}]{atkins1959ions}%
  \BibitemOpen
  \bibfield  {author} {\bibinfo {author} {\bibfnamefont {K.}~\bibnamefont
  {Atkins}},\ }\bibfield  {title} {\enquote {\bibinfo {title} {Ions in liquid
  helium},}\ }\href@noop {} {\bibfield  {journal} {\bibinfo  {journal}
  {Physical Review}\ }\textbf {\bibinfo {volume} {116}},\ \bibinfo {pages}
  {1339} (\bibinfo {year} {1959})}\BibitemShut {NoStop}%
\bibitem [{\citenamefont {M{\"u}ller}, \citenamefont {Mudrich},\ and\
  \citenamefont {Stienkemeier}(2009)}]{muller2009alkali}%
  \BibitemOpen
  \bibfield  {author} {\bibinfo {author} {\bibfnamefont {S.}~\bibnamefont
  {M{\"u}ller}}, \bibinfo {author} {\bibfnamefont {M.}~\bibnamefont {Mudrich}},
  \ and\ \bibinfo {author} {\bibfnamefont {F.}~\bibnamefont {Stienkemeier}},\
  }\bibfield  {title} {\enquote {\bibinfo {title} {Alkali-helium snowball
  complexes formed on helium nanodroplets},}\ }\href@noop {} {\bibfield
  {journal} {\bibinfo  {journal} {J. Chem. Phys.}\ }\textbf {\bibinfo {volume}
  {131}},\ \bibinfo {pages} {044319} (\bibinfo {year} {2009})}\BibitemShut
  {NoStop}%
\bibitem [{\citenamefont {Theisen}, \citenamefont {Lackner},\ and\
  \citenamefont {Ernst}(2010)}]{theisen2010forming}%
  \BibitemOpen
  \bibfield  {author} {\bibinfo {author} {\bibfnamefont {M.}~\bibnamefont
  {Theisen}}, \bibinfo {author} {\bibfnamefont {F.}~\bibnamefont {Lackner}}, \
  and\ \bibinfo {author} {\bibfnamefont {W.~E.}\ \bibnamefont {Ernst}},\
  }\bibfield  {title} {\enquote {\bibinfo {title} {Forming rb+ snowballs in the
  center of he nanodroplets},}\ }\href@noop {} {\bibfield  {journal} {\bibinfo
  {journal} {Phys. Chem. Chem. Phys.}\ }\textbf {\bibinfo {volume} {12}},\
  \bibinfo {pages} {14861--14863} (\bibinfo {year} {2010})}\BibitemShut
  {NoStop}%
\bibitem [{\citenamefont {Harms}, \citenamefont {Toennies},\ and\ \citenamefont
  {Dalfovo}(1998)}]{harms1998density}%
  \BibitemOpen
  \bibfield  {author} {\bibinfo {author} {\bibfnamefont {J.}~\bibnamefont
  {Harms}}, \bibinfo {author} {\bibfnamefont {J.~P.}\ \bibnamefont {Toennies}},
  \ and\ \bibinfo {author} {\bibfnamefont {F.}~\bibnamefont {Dalfovo}},\
  }\bibfield  {title} {\enquote {\bibinfo {title} {Density of superfluid helium
  droplets},}\ }\href@noop {} {\bibfield  {journal} {\bibinfo  {journal}
  {Physical Review B}\ }\textbf {\bibinfo {volume} {58}},\ \bibinfo {pages}
  {3341} (\bibinfo {year} {1998})}\BibitemShut {NoStop}%
\bibitem [{\citenamefont {Zhang}\ and\ \citenamefont
  {Drabbels}(2012)}]{zhang2012communication}%
  \BibitemOpen
  \bibfield  {author} {\bibinfo {author} {\bibfnamefont {X.}~\bibnamefont
  {Zhang}}\ and\ \bibinfo {author} {\bibfnamefont {M.}~\bibnamefont
  {Drabbels}},\ }\bibfield  {title} {\enquote {\bibinfo {title} {Communication:
  {B}arium ions and helium nanodroplets: {S}olvation and desolvation},}\
  }\href@noop {} {\bibfield  {journal} {\bibinfo  {journal} {J. Chem. Phys.}\
  }\textbf {\bibinfo {volume} {137}},\ \bibinfo {pages} {051102} (\bibinfo
  {year} {2012})}\BibitemShut {NoStop}%
\bibitem [{\citenamefont {Loginov}\ and\ \citenamefont
  {Drabbels}(2012)}]{loginov2012spectroscopy}%
  \BibitemOpen
  \bibfield  {author} {\bibinfo {author} {\bibfnamefont {E.}~\bibnamefont
  {Loginov}}\ and\ \bibinfo {author} {\bibfnamefont {M.}~\bibnamefont
  {Drabbels}},\ }\bibfield  {title} {\enquote {\bibinfo {title} {Spectroscopy
  and dynamics of barium-doped helium nanodroplets},}\ }\href@noop {}
  {\bibfield  {journal} {\bibinfo  {journal} {J. Chem. Phys.}\ }\textbf
  {\bibinfo {volume} {136}},\ \bibinfo {pages} {154302} (\bibinfo {year}
  {2012})}\BibitemShut {NoStop}%
\bibitem [{\citenamefont {von Vangerow}\ \emph {et~al.}(2015)\citenamefont {von
  Vangerow}, \citenamefont {John}, \citenamefont {Stienkemeier},\ and\
  \citenamefont {Mudrich}}]{von2015dynamics}%
  \BibitemOpen
  \bibfield  {author} {\bibinfo {author} {\bibfnamefont {J.}~\bibnamefont {von
  Vangerow}}, \bibinfo {author} {\bibfnamefont {O.}~\bibnamefont {John}},
  \bibinfo {author} {\bibfnamefont {F.}~\bibnamefont {Stienkemeier}}, \ and\
  \bibinfo {author} {\bibfnamefont {M.}~\bibnamefont {Mudrich}},\ }\bibfield
  {title} {\enquote {\bibinfo {title} {Dynamics of solvation and desolvation of
  rubidium attached to {H}e nanodroplets},}\ }\href@noop {} {\bibfield
  {journal} {\bibinfo  {journal} {J. Chem. Phys.}\ }\textbf {\bibinfo {volume}
  {143}},\ \bibinfo {pages} {034302} (\bibinfo {year} {2015})}\BibitemShut
  {NoStop}%
\bibitem [{\citenamefont {Pickering}\ \emph {et~al.}(2018)\citenamefont
  {Pickering}, \citenamefont {Shepperson}, \citenamefont {Christiansen},\ and\
  \citenamefont {Stapelfeldt}}]{pickering2018femtosecond}%
  \BibitemOpen
  \bibfield  {author} {\bibinfo {author} {\bibfnamefont {J.~D.}\ \bibnamefont
  {Pickering}}, \bibinfo {author} {\bibfnamefont {B.}~\bibnamefont
  {Shepperson}}, \bibinfo {author} {\bibfnamefont {L.}~\bibnamefont
  {Christiansen}}, \ and\ \bibinfo {author} {\bibfnamefont {H.}~\bibnamefont
  {Stapelfeldt}},\ }\bibfield  {title} {\enquote {\bibinfo {title} {Femtosecond
  laser induced coulomb explosion imaging of aligned {OCS} oligomers inside
  helium nanodroplets},}\ }\href@noop {} {\bibfield  {journal} {\bibinfo
  {journal} {J. Chem. Phys.}\ }\textbf {\bibinfo {volume} {149}},\ \bibinfo
  {pages} {154306} (\bibinfo {year} {2018})}\BibitemShut {NoStop}%
\bibitem [{\citenamefont {F{\'a}rn{\'\i}k}\ \emph {et~al.}(1998)\citenamefont
  {F{\'a}rn{\'\i}k}, \citenamefont {Henne}, \citenamefont {Samelin},\ and\
  \citenamefont {Toennies}}]{farnik1998differences}%
  \BibitemOpen
  \bibfield  {author} {\bibinfo {author} {\bibfnamefont {M.}~\bibnamefont
  {F{\'a}rn{\'\i}k}}, \bibinfo {author} {\bibfnamefont {U.}~\bibnamefont
  {Henne}}, \bibinfo {author} {\bibfnamefont {B.}~\bibnamefont {Samelin}}, \
  and\ \bibinfo {author} {\bibfnamefont {J.~P.}\ \bibnamefont {Toennies}},\
  }\bibfield  {title} {\enquote {\bibinfo {title} {Differences in the
  detachment of electron bubbles from superfluid 4{H}e droplets versus
  nonsuperfluid 3{H}e droplets},}\ }\href@noop {} {\bibfield  {journal}
  {\bibinfo  {journal} {Phys. Rev. Lett.}\ }\textbf {\bibinfo {volume} {81}},\
  \bibinfo {pages} {3892} (\bibinfo {year} {1998})}\BibitemShut {NoStop}%
\bibitem [{\citenamefont {Rosenblit}\ and\ \citenamefont
  {Jortner}(2006)}]{rosenblit2006electron}%
  \BibitemOpen
  \bibfield  {author} {\bibinfo {author} {\bibfnamefont {M.}~\bibnamefont
  {Rosenblit}}\ and\ \bibinfo {author} {\bibfnamefont {J.}~\bibnamefont
  {Jortner}},\ }\bibfield  {title} {\enquote {\bibinfo {title} {Electron
  bubbles in helium clusters. {I}. {S}tructure and energetics},}\ }\href@noop
  {} {\bibfield  {journal} {\bibinfo  {journal} {J. Chem. Phys.}\ }\textbf
  {\bibinfo {volume} {124}},\ \bibinfo {pages} {064502} (\bibinfo {year}
  {2006})}\BibitemShut {NoStop}%
\bibitem [{\citenamefont {Buchta}\ \emph
  {et~al.}(2013{\natexlab{a}})\citenamefont {Buchta}, \citenamefont {Krishnan},
  \citenamefont {Brauer}, \citenamefont {Drabbels}, \citenamefont {O’Keeffe},
  \citenamefont {Devetta}, \citenamefont {Di~Fraia}, \citenamefont {Callegari},
  \citenamefont {Richter}, \citenamefont {Coreno} \emph
  {et~al.}}]{buchta2013extreme}%
  \BibitemOpen
  \bibfield  {author} {\bibinfo {author} {\bibfnamefont {D.}~\bibnamefont
  {Buchta}}, \bibinfo {author} {\bibfnamefont {S.~R.}\ \bibnamefont
  {Krishnan}}, \bibinfo {author} {\bibfnamefont {N.~B.}\ \bibnamefont
  {Brauer}}, \bibinfo {author} {\bibfnamefont {M.}~\bibnamefont {Drabbels}},
  \bibinfo {author} {\bibfnamefont {P.}~\bibnamefont {O’Keeffe}}, \bibinfo
  {author} {\bibfnamefont {M.}~\bibnamefont {Devetta}}, \bibinfo {author}
  {\bibfnamefont {M.}~\bibnamefont {Di~Fraia}}, \bibinfo {author}
  {\bibfnamefont {C.}~\bibnamefont {Callegari}}, \bibinfo {author}
  {\bibfnamefont {R.}~\bibnamefont {Richter}}, \bibinfo {author} {\bibfnamefont
  {M.}~\bibnamefont {Coreno}},  \emph {et~al.},\ }\bibfield  {title} {\enquote
  {\bibinfo {title} {Extreme ultraviolet ionization of pure he nanodroplets:
  Mass-correlated photoelectron imaging, penning ionization, and electron
  energy-loss spectra},}\ }\href@noop {} {\bibfield  {journal} {\bibinfo
  {journal} {J. Chem. Phys.}\ }\textbf {\bibinfo {volume} {139}},\ \bibinfo
  {pages} {084301} (\bibinfo {year} {2013}{\natexlab{a}})}\BibitemShut
  {NoStop}%
\bibitem [{\citenamefont {Buchta}\ \emph
  {et~al.}(2013{\natexlab{b}})\citenamefont {Buchta}, \citenamefont {Krishnan},
  \citenamefont {Brauer}, \citenamefont {Drabbels}, \citenamefont {O’Keeffe},
  \citenamefont {Devetta}, \citenamefont {Di~Fraia}, \citenamefont {Callegari},
  \citenamefont {Richter}, \citenamefont {Coreno} \emph
  {et~al.}}]{buchta2013charge}%
  \BibitemOpen
  \bibfield  {author} {\bibinfo {author} {\bibfnamefont {D.}~\bibnamefont
  {Buchta}}, \bibinfo {author} {\bibfnamefont {S.~R.}\ \bibnamefont
  {Krishnan}}, \bibinfo {author} {\bibfnamefont {N.~B.}\ \bibnamefont
  {Brauer}}, \bibinfo {author} {\bibfnamefont {M.}~\bibnamefont {Drabbels}},
  \bibinfo {author} {\bibfnamefont {P.}~\bibnamefont {O’Keeffe}}, \bibinfo
  {author} {\bibfnamefont {M.}~\bibnamefont {Devetta}}, \bibinfo {author}
  {\bibfnamefont {M.}~\bibnamefont {Di~Fraia}}, \bibinfo {author}
  {\bibfnamefont {C.}~\bibnamefont {Callegari}}, \bibinfo {author}
  {\bibfnamefont {R.}~\bibnamefont {Richter}}, \bibinfo {author} {\bibfnamefont
  {M.}~\bibnamefont {Coreno}},  \emph {et~al.},\ }\bibfield  {title} {\enquote
  {\bibinfo {title} {Charge transfer and penning ionization of dopants in or on
  helium nanodroplets exposed to {EUV} radiation},}\ }\href@noop {} {\bibfield
  {journal} {\bibinfo  {journal} {J. Phys. Chem. A}\ }\textbf {\bibinfo
  {volume} {117}},\ \bibinfo {pages} {4394--4403} (\bibinfo {year}
  {2013}{\natexlab{b}})}\BibitemShut {NoStop}%
\bibitem [{\citenamefont {LaForge}\ \emph {et~al.}(2016)\citenamefont
  {LaForge}, \citenamefont {Stumpf}, \citenamefont {Gokhberg}, \citenamefont
  {von Vangerow}, \citenamefont {Stienkemeier}, \citenamefont {Kryzhevoi},
  \citenamefont {O’Keeffe}, \citenamefont {Ciavardini}, \citenamefont
  {Krishnan}, \citenamefont {Coreno} \emph {et~al.}}]{laforge2016enhanced}%
  \BibitemOpen
  \bibfield  {author} {\bibinfo {author} {\bibfnamefont {A.}~\bibnamefont
  {LaForge}}, \bibinfo {author} {\bibfnamefont {V.}~\bibnamefont {Stumpf}},
  \bibinfo {author} {\bibfnamefont {K.}~\bibnamefont {Gokhberg}}, \bibinfo
  {author} {\bibfnamefont {J.}~\bibnamefont {von Vangerow}}, \bibinfo {author}
  {\bibfnamefont {F.}~\bibnamefont {Stienkemeier}}, \bibinfo {author}
  {\bibfnamefont {N.}~\bibnamefont {Kryzhevoi}}, \bibinfo {author}
  {\bibfnamefont {P.}~\bibnamefont {O’Keeffe}}, \bibinfo {author}
  {\bibfnamefont {A.}~\bibnamefont {Ciavardini}}, \bibinfo {author}
  {\bibfnamefont {S.}~\bibnamefont {Krishnan}}, \bibinfo {author}
  {\bibfnamefont {M.}~\bibnamefont {Coreno}},  \emph {et~al.},\ }\bibfield
  {title} {\enquote {\bibinfo {title} {Enhanced ionization of embedded clusters
  by electron-transfer-mediated decay in helium nanodroplets},}\ }\href@noop {}
  {\bibfield  {journal} {\bibinfo  {journal} {Phys. Rev. Lett.}\ }\textbf
  {\bibinfo {volume} {116}},\ \bibinfo {pages} {203001} (\bibinfo {year}
  {2016})}\BibitemShut {NoStop}%
\bibitem [{\citenamefont {Shcherbinin}\ \emph {et~al.}(2017)\citenamefont
  {Shcherbinin}, \citenamefont {LaForge}, \citenamefont {Sharma}, \citenamefont
  {Devetta}, \citenamefont {Richter}, \citenamefont {Moshammer}, \citenamefont
  {Pfeifer},\ and\ \citenamefont {Mudrich}}]{shcherbinin2017interatomic}%
  \BibitemOpen
  \bibfield  {author} {\bibinfo {author} {\bibfnamefont {M.}~\bibnamefont
  {Shcherbinin}}, \bibinfo {author} {\bibfnamefont {A.}~\bibnamefont
  {LaForge}}, \bibinfo {author} {\bibfnamefont {V.}~\bibnamefont {Sharma}},
  \bibinfo {author} {\bibfnamefont {M.}~\bibnamefont {Devetta}}, \bibinfo
  {author} {\bibfnamefont {R.}~\bibnamefont {Richter}}, \bibinfo {author}
  {\bibfnamefont {R.}~\bibnamefont {Moshammer}}, \bibinfo {author}
  {\bibfnamefont {T.}~\bibnamefont {Pfeifer}}, \ and\ \bibinfo {author}
  {\bibfnamefont {M.}~\bibnamefont {Mudrich}},\ }\bibfield  {title} {\enquote
  {\bibinfo {title} {Interatomic coulombic decay in helium nanodroplets},}\
  }\href@noop {} {\bibfield  {journal} {\bibinfo  {journal} {Physical Review
  A}\ }\textbf {\bibinfo {volume} {96}},\ \bibinfo {pages} {013407} (\bibinfo
  {year} {2017})}\BibitemShut {NoStop}%
\bibitem [{\citenamefont {Shcherbinin}\ \emph {et~al.}(2019)\citenamefont
  {Shcherbinin}, \citenamefont {Westergaard}, \citenamefont {Hanif},
  \citenamefont {Krishnan}, \citenamefont {LaForge}, \citenamefont {Richter},
  \citenamefont {Pfeifer},\ and\ \citenamefont
  {Mudrich}}]{shcherbinin2019inelastic}%
  \BibitemOpen
  \bibfield  {author} {\bibinfo {author} {\bibfnamefont {M.}~\bibnamefont
  {Shcherbinin}}, \bibinfo {author} {\bibfnamefont {F.~V.}\ \bibnamefont
  {Westergaard}}, \bibinfo {author} {\bibfnamefont {M.}~\bibnamefont {Hanif}},
  \bibinfo {author} {\bibfnamefont {S.}~\bibnamefont {Krishnan}}, \bibinfo
  {author} {\bibfnamefont {A.}~\bibnamefont {LaForge}}, \bibinfo {author}
  {\bibfnamefont {R.}~\bibnamefont {Richter}}, \bibinfo {author} {\bibfnamefont
  {T.}~\bibnamefont {Pfeifer}}, \ and\ \bibinfo {author} {\bibfnamefont
  {M.}~\bibnamefont {Mudrich}},\ }\bibfield  {title} {\enquote {\bibinfo
  {title} {Inelastic scattering of photoelectrons from he nanodroplets},}\
  }\href@noop {} {\bibfield  {journal} {\bibinfo  {journal} {J. Chem. Phys.}\
  }\textbf {\bibinfo {volume} {150}},\ \bibinfo {pages} {044304} (\bibinfo
  {year} {2019})}\BibitemShut {NoStop}%
\bibitem [{\citenamefont {Ben~Ltaief}\ \emph {et~al.}(2019)\citenamefont
  {Ben~Ltaief}, \citenamefont {Shcherbinin}, \citenamefont {Mandal},
  \citenamefont {Krishnan}, \citenamefont {LaForge}, \citenamefont {Richter},
  \citenamefont {Turchini}, \citenamefont {Zema}, \citenamefont {Pfeifer},
  \citenamefont {Fasshauer} \emph {et~al.}}]{ben2019charge}%
  \BibitemOpen
  \bibfield  {author} {\bibinfo {author} {\bibfnamefont {L.}~\bibnamefont
  {Ben~Ltaief}}, \bibinfo {author} {\bibfnamefont {M.}~\bibnamefont
  {Shcherbinin}}, \bibinfo {author} {\bibfnamefont {S.}~\bibnamefont {Mandal}},
  \bibinfo {author} {\bibfnamefont {S.}~\bibnamefont {Krishnan}}, \bibinfo
  {author} {\bibfnamefont {A.}~\bibnamefont {LaForge}}, \bibinfo {author}
  {\bibfnamefont {R.}~\bibnamefont {Richter}}, \bibinfo {author} {\bibfnamefont
  {S.}~\bibnamefont {Turchini}}, \bibinfo {author} {\bibfnamefont
  {N.}~\bibnamefont {Zema}}, \bibinfo {author} {\bibfnamefont {T.}~\bibnamefont
  {Pfeifer}}, \bibinfo {author} {\bibfnamefont {E.}~\bibnamefont {Fasshauer}},
  \emph {et~al.},\ }\bibfield  {title} {\enquote {\bibinfo {title} {Charge
  exchange dominates long-range interatomic coulombic decay of excited
  metal-doped helium nanodroplets},}\ }\href@noop {} {\bibfield  {journal}
  {\bibinfo  {journal} {J. Phys. Chem. Lett.}\ }\textbf {\bibinfo {volume}
  {10}},\ \bibinfo {pages} {6904--6909} (\bibinfo {year} {2019})}\BibitemShut
  {NoStop}%
\bibitem [{\citenamefont {LaForge}\ \emph {et~al.}(2019)\citenamefont
  {LaForge}, \citenamefont {Shcherbinin}, \citenamefont {Stienkemeier},
  \citenamefont {Richter}, \citenamefont {Moshammer}, \citenamefont {Pfeifer},\
  and\ \citenamefont {Mudrich}}]{laforge2019highly}%
  \BibitemOpen
  \bibfield  {author} {\bibinfo {author} {\bibfnamefont {A.}~\bibnamefont
  {LaForge}}, \bibinfo {author} {\bibfnamefont {M.}~\bibnamefont
  {Shcherbinin}}, \bibinfo {author} {\bibfnamefont {F.}~\bibnamefont
  {Stienkemeier}}, \bibinfo {author} {\bibfnamefont {R.}~\bibnamefont
  {Richter}}, \bibinfo {author} {\bibfnamefont {R.}~\bibnamefont {Moshammer}},
  \bibinfo {author} {\bibfnamefont {T.}~\bibnamefont {Pfeifer}}, \ and\
  \bibinfo {author} {\bibfnamefont {M.}~\bibnamefont {Mudrich}},\ }\bibfield
  {title} {\enquote {\bibinfo {title} {Highly efficient double ionization of
  mixed alkali dimers by intermolecular coulombic decay},}\ }\href@noop {}
  {\bibfield  {journal} {\bibinfo  {journal} {Nature Physics}\ }\textbf
  {\bibinfo {volume} {15}},\ \bibinfo {pages} {247--250} (\bibinfo {year}
  {2019})}\BibitemShut {NoStop}%
\bibitem [{\citenamefont {Ltaief}\ \emph {et~al.}(2020)\citenamefont {Ltaief},
  \citenamefont {Shcherbinin}, \citenamefont {Mandal}, \citenamefont
  {Krishnan}, \citenamefont {Richter}, \citenamefont {Pfeifer}, \citenamefont
  {Bauer}, \citenamefont {Ghosh}, \citenamefont {Mudrich}, \citenamefont
  {Gokhberg} \emph {et~al.}}]{ltaief2020electron}%
  \BibitemOpen
  \bibfield  {author} {\bibinfo {author} {\bibfnamefont {L.~B.}\ \bibnamefont
  {Ltaief}}, \bibinfo {author} {\bibfnamefont {M.}~\bibnamefont {Shcherbinin}},
  \bibinfo {author} {\bibfnamefont {S.}~\bibnamefont {Mandal}}, \bibinfo
  {author} {\bibfnamefont {S.}~\bibnamefont {Krishnan}}, \bibinfo {author}
  {\bibfnamefont {R.}~\bibnamefont {Richter}}, \bibinfo {author} {\bibfnamefont
  {T.}~\bibnamefont {Pfeifer}}, \bibinfo {author} {\bibfnamefont
  {M.}~\bibnamefont {Bauer}}, \bibinfo {author} {\bibfnamefont
  {A.}~\bibnamefont {Ghosh}}, \bibinfo {author} {\bibfnamefont
  {M.}~\bibnamefont {Mudrich}}, \bibinfo {author} {\bibfnamefont
  {K.}~\bibnamefont {Gokhberg}},  \emph {et~al.},\ }\bibfield  {title}
  {\enquote {\bibinfo {title} {Electron transfer mediated decay of alkali
  dimers attached to he nanodroplets},}\ }\href@noop {} {\bibfield  {journal}
  {\bibinfo  {journal} {Phys. Chem. Chem. Phys.}\ }\textbf {\bibinfo {volume}
  {22}},\ \bibinfo {pages} {8557--8564} (\bibinfo {year} {2020})}\BibitemShut
  {NoStop}%
\bibitem [{\citenamefont {Jarvis}\ \emph {et~al.}(1999)\citenamefont {Jarvis},
  \citenamefont {Weitzel}, \citenamefont {Malow}, \citenamefont {Baer},
  \citenamefont {Song},\ and\ \citenamefont {Ng}}]{jarvis1999high}%
  \BibitemOpen
  \bibfield  {author} {\bibinfo {author} {\bibfnamefont {G.}~\bibnamefont
  {Jarvis}}, \bibinfo {author} {\bibfnamefont {K.-M.}\ \bibnamefont {Weitzel}},
  \bibinfo {author} {\bibfnamefont {M.}~\bibnamefont {Malow}}, \bibinfo
  {author} {\bibfnamefont {T.}~\bibnamefont {Baer}}, \bibinfo {author}
  {\bibfnamefont {Y.}~\bibnamefont {Song}}, \ and\ \bibinfo {author}
  {\bibfnamefont {C.}~\bibnamefont {Ng}},\ }\bibfield  {title} {\enquote
  {\bibinfo {title} {High-resolution pulsed field ionization
  photoelectron-photoion coincidence spectroscopy using synchrotron
  radiation},}\ }\href@noop {} {\bibfield  {journal} {\bibinfo  {journal} {Rev.
  Sci. Instrum.}\ }\textbf {\bibinfo {volume} {70}},\ \bibinfo {pages}
  {3892--3906} (\bibinfo {year} {1999})}\BibitemShut {NoStop}%
\bibitem [{\citenamefont {Ueda}\ and\ \citenamefont
  {Eland}(2005)}]{ueda2005molecular}%
  \BibitemOpen
  \bibfield  {author} {\bibinfo {author} {\bibfnamefont {K.}~\bibnamefont
  {Ueda}}\ and\ \bibinfo {author} {\bibfnamefont {J.~H.}\ \bibnamefont
  {Eland}},\ }\bibfield  {title} {\enquote {\bibinfo {title} {Molecular
  photodissociation studied by {VUV} and soft {X}-ray radiation},}\ }\href@noop
  {} {\bibfield  {journal} {\bibinfo  {journal} {J. Phys. B: At., Mol. Opt.
  Phys.}\ }\textbf {\bibinfo {volume} {38}},\ \bibinfo {pages} {S839} (\bibinfo
  {year} {2005})}\BibitemShut {NoStop}%
\bibitem [{\citenamefont {Bastian}\ \emph {et~al.}(2022)\citenamefont
  {Bastian}, \citenamefont {Asmussen}, \citenamefont {Ben~Ltaief},
  \citenamefont {Czasch}, \citenamefont {Jones}, \citenamefont {Hoffmann},
  \citenamefont {Pedersen},\ and\ \citenamefont {Mudrich}}]{bastian2022new}%
  \BibitemOpen
  \bibfield  {author} {\bibinfo {author} {\bibfnamefont {B.}~\bibnamefont
  {Bastian}}, \bibinfo {author} {\bibfnamefont {J.~D.}\ \bibnamefont
  {Asmussen}}, \bibinfo {author} {\bibfnamefont {L.}~\bibnamefont
  {Ben~Ltaief}}, \bibinfo {author} {\bibfnamefont {A.}~\bibnamefont {Czasch}},
  \bibinfo {author} {\bibfnamefont {N.~C.}\ \bibnamefont {Jones}}, \bibinfo
  {author} {\bibfnamefont {S.~V.}\ \bibnamefont {Hoffmann}}, \bibinfo {author}
  {\bibfnamefont {H.~B.}\ \bibnamefont {Pedersen}}, \ and\ \bibinfo {author}
  {\bibfnamefont {M.}~\bibnamefont {Mudrich}},\ }\bibfield  {title} {\enquote
  {\bibinfo {title} {A new endstation for extreme-ultraviolet spectroscopy of
  free clusters and nanodroplets},}\ }\href@noop {} {\bibfield  {journal}
  {\bibinfo  {journal} {Rev. Sci. Instrum.}\ }\textbf {\bibinfo {volume}
  {93}},\ \bibinfo {pages} {075110} (\bibinfo {year} {2022})}\BibitemShut
  {NoStop}%
\bibitem [{\citenamefont {Mudrich}\ \emph {et~al.}(2020)\citenamefont
  {Mudrich}, \citenamefont {LaForge}, \citenamefont {Ciavardini}, \citenamefont
  {O’Keeffe}, \citenamefont {Callegari}, \citenamefont {Coreno},
  \citenamefont {Demidovich}, \citenamefont {Devetta}, \citenamefont {Fraia},
  \citenamefont {Drabbels} \emph {et~al.}}]{mudrich2020ultrafast}%
  \BibitemOpen
  \bibfield  {author} {\bibinfo {author} {\bibfnamefont {M.}~\bibnamefont
  {Mudrich}}, \bibinfo {author} {\bibfnamefont {A.}~\bibnamefont {LaForge}},
  \bibinfo {author} {\bibfnamefont {A.}~\bibnamefont {Ciavardini}}, \bibinfo
  {author} {\bibfnamefont {P.}~\bibnamefont {O’Keeffe}}, \bibinfo {author}
  {\bibfnamefont {C.}~\bibnamefont {Callegari}}, \bibinfo {author}
  {\bibfnamefont {M.}~\bibnamefont {Coreno}}, \bibinfo {author} {\bibfnamefont
  {A.}~\bibnamefont {Demidovich}}, \bibinfo {author} {\bibfnamefont
  {M.}~\bibnamefont {Devetta}}, \bibinfo {author} {\bibfnamefont {M.~D.}\
  \bibnamefont {Fraia}}, \bibinfo {author} {\bibfnamefont {M.}~\bibnamefont
  {Drabbels}},  \emph {et~al.},\ }\bibfield  {title} {\enquote {\bibinfo
  {title} {Ultrafast relaxation of photoexcited superfluid he nanodroplets},}\
  }\href@noop {} {\bibfield  {journal} {\bibinfo  {journal} {Nature
  communications}\ }\textbf {\bibinfo {volume} {11}},\ \bibinfo {pages} {112}
  (\bibinfo {year} {2020})}\BibitemShut {NoStop}%
\bibitem [{\citenamefont {LaForge}\ \emph {et~al.}(2022)\citenamefont
  {LaForge}, \citenamefont {Asmussen}, \citenamefont {Bastian}, \citenamefont
  {Bonanomi}, \citenamefont {Callegari}, \citenamefont {De}, \citenamefont
  {Di~Fraia}, \citenamefont {Gorman}, \citenamefont {Hartweg}, \citenamefont
  {Krishnan} \emph {et~al.}}]{laforge2022relaxation}%
  \BibitemOpen
  \bibfield  {author} {\bibinfo {author} {\bibfnamefont {A.}~\bibnamefont
  {LaForge}}, \bibinfo {author} {\bibfnamefont {J.}~\bibnamefont {Asmussen}},
  \bibinfo {author} {\bibfnamefont {B.}~\bibnamefont {Bastian}}, \bibinfo
  {author} {\bibfnamefont {M.}~\bibnamefont {Bonanomi}}, \bibinfo {author}
  {\bibfnamefont {C.}~\bibnamefont {Callegari}}, \bibinfo {author}
  {\bibfnamefont {S.}~\bibnamefont {De}}, \bibinfo {author} {\bibfnamefont
  {M.}~\bibnamefont {Di~Fraia}}, \bibinfo {author} {\bibfnamefont
  {L.}~\bibnamefont {Gorman}}, \bibinfo {author} {\bibfnamefont
  {S.}~\bibnamefont {Hartweg}}, \bibinfo {author} {\bibfnamefont
  {S.}~\bibnamefont {Krishnan}},  \emph {et~al.},\ }\bibfield  {title}
  {\enquote {\bibinfo {title} {Relaxation dynamics in excited helium
  nanodroplets probed with high resolution, time-resolved photoelectron
  spectroscopy},}\ }\href@noop {} {\bibfield  {journal} {\bibinfo  {journal}
  {Phys. Chem. Chem. Phys.}\ }\textbf {\bibinfo {volume} {24}},\ \bibinfo
  {pages} {28844--28852} (\bibinfo {year} {2022})}\BibitemShut {NoStop}%
\bibitem [{\citenamefont {Averbukh}, \citenamefont {M{\"u}ller},\ and\
  \citenamefont {Cederbaum}(2004)}]{averbukh2004mechanism}%
  \BibitemOpen
  \bibfield  {author} {\bibinfo {author} {\bibfnamefont {V.}~\bibnamefont
  {Averbukh}}, \bibinfo {author} {\bibfnamefont {I.~B.}\ \bibnamefont
  {M{\"u}ller}}, \ and\ \bibinfo {author} {\bibfnamefont {L.~S.}\ \bibnamefont
  {Cederbaum}},\ }\bibfield  {title} {\enquote {\bibinfo {title} {Mechanism of
  interatomic coulombic decay in clusters},}\ }\href@noop {} {\bibfield
  {journal} {\bibinfo  {journal} {Phys. Rev. Lett.}\ }\textbf {\bibinfo
  {volume} {93}},\ \bibinfo {pages} {263002} (\bibinfo {year}
  {2004})}\BibitemShut {NoStop}%
\bibitem [{\citenamefont {Kuleff}\ \emph {et~al.}(2010)\citenamefont {Kuleff},
  \citenamefont {Gokhberg}, \citenamefont {Kopelke},\ and\ \citenamefont
  {Cederbaum}}]{kuleff2010ultrafast}%
  \BibitemOpen
  \bibfield  {author} {\bibinfo {author} {\bibfnamefont {A.~I.}\ \bibnamefont
  {Kuleff}}, \bibinfo {author} {\bibfnamefont {K.}~\bibnamefont {Gokhberg}},
  \bibinfo {author} {\bibfnamefont {S.}~\bibnamefont {Kopelke}}, \ and\
  \bibinfo {author} {\bibfnamefont {L.~S.}\ \bibnamefont {Cederbaum}},\
  }\bibfield  {title} {\enquote {\bibinfo {title} {Ultrafast interatomic
  electronic decay in multiply excited clusters},}\ }\href@noop {} {\bibfield
  {journal} {\bibinfo  {journal} {Phys. Rev. Lett.}\ }\textbf {\bibinfo
  {volume} {105}},\ \bibinfo {pages} {043004} (\bibinfo {year}
  {2010})}\BibitemShut {NoStop}%
\bibitem [{\citenamefont {Peterka}\ \emph {et~al.}(2006)\citenamefont
  {Peterka}, \citenamefont {Kim}, \citenamefont {Wang},\ and\ \citenamefont
  {Neumark}}]{peterka2006photoionization}%
  \BibitemOpen
  \bibfield  {author} {\bibinfo {author} {\bibfnamefont {D.~S.}\ \bibnamefont
  {Peterka}}, \bibinfo {author} {\bibfnamefont {J.~H.}\ \bibnamefont {Kim}},
  \bibinfo {author} {\bibfnamefont {C.~C.}\ \bibnamefont {Wang}}, \ and\
  \bibinfo {author} {\bibfnamefont {D.~M.}\ \bibnamefont {Neumark}},\
  }\bibfield  {title} {\enquote {\bibinfo {title} {Photoionization and
  photofragmentation of sf6 in helium nanodroplets},}\ }\href@noop {}
  {\bibfield  {journal} {\bibinfo  {journal} {The Journal of Physical Chemistry
  B}\ }\textbf {\bibinfo {volume} {110}},\ \bibinfo {pages} {19945--19955}
  (\bibinfo {year} {2006})}\BibitemShut {NoStop}%
\bibitem [{\citenamefont {Ovcharenko}\ \emph {et~al.}(2014)\citenamefont
  {Ovcharenko}, \citenamefont {Lyamayev}, \citenamefont {Katzy}, \citenamefont
  {Devetta}, \citenamefont {LaForge}, \citenamefont {O’Keeffe}, \citenamefont
  {Plekan}, \citenamefont {Finetti}, \citenamefont {Di~Fraia}, \citenamefont
  {Mudrich} \emph {et~al.}}]{ovcharenko2014novel}%
  \BibitemOpen
  \bibfield  {author} {\bibinfo {author} {\bibfnamefont {Y.}~\bibnamefont
  {Ovcharenko}}, \bibinfo {author} {\bibfnamefont {V.}~\bibnamefont
  {Lyamayev}}, \bibinfo {author} {\bibfnamefont {R.}~\bibnamefont {Katzy}},
  \bibinfo {author} {\bibfnamefont {M.}~\bibnamefont {Devetta}}, \bibinfo
  {author} {\bibfnamefont {A.}~\bibnamefont {LaForge}}, \bibinfo {author}
  {\bibfnamefont {P.}~\bibnamefont {O’Keeffe}}, \bibinfo {author}
  {\bibfnamefont {O.}~\bibnamefont {Plekan}}, \bibinfo {author} {\bibfnamefont
  {P.}~\bibnamefont {Finetti}}, \bibinfo {author} {\bibfnamefont
  {M.}~\bibnamefont {Di~Fraia}}, \bibinfo {author} {\bibfnamefont
  {M.}~\bibnamefont {Mudrich}},  \emph {et~al.},\ }\bibfield  {title} {\enquote
  {\bibinfo {title} {Novel collective autoionization process observed in
  electron spectra of he clusters},}\ }\href@noop {} {\bibfield  {journal}
  {\bibinfo  {journal} {Phys. Rev. Lett.}\ }\textbf {\bibinfo {volume} {112}},\
  \bibinfo {pages} {073401} (\bibinfo {year} {2014})}\BibitemShut {NoStop}%
\bibitem [{\citenamefont {LaForge}\ \emph {et~al.}(2014)\citenamefont
  {LaForge}, \citenamefont {Drabbels}, \citenamefont {Brauer}, \citenamefont
  {Coreno}, \citenamefont {Devetta}, \citenamefont {Di~Fraia}, \citenamefont
  {Finetti}, \citenamefont {Grazioli}, \citenamefont {Katzy}, \citenamefont
  {Lyamayev} \emph {et~al.}}]{laforge2014collective}%
  \BibitemOpen
  \bibfield  {author} {\bibinfo {author} {\bibfnamefont {A.}~\bibnamefont
  {LaForge}}, \bibinfo {author} {\bibfnamefont {M.}~\bibnamefont {Drabbels}},
  \bibinfo {author} {\bibfnamefont {N.~B.}\ \bibnamefont {Brauer}}, \bibinfo
  {author} {\bibfnamefont {M.}~\bibnamefont {Coreno}}, \bibinfo {author}
  {\bibfnamefont {M.}~\bibnamefont {Devetta}}, \bibinfo {author} {\bibfnamefont
  {M.}~\bibnamefont {Di~Fraia}}, \bibinfo {author} {\bibfnamefont
  {P.}~\bibnamefont {Finetti}}, \bibinfo {author} {\bibfnamefont
  {C.}~\bibnamefont {Grazioli}}, \bibinfo {author} {\bibfnamefont
  {R.}~\bibnamefont {Katzy}}, \bibinfo {author} {\bibfnamefont
  {V.}~\bibnamefont {Lyamayev}},  \emph {et~al.},\ }\bibfield  {title}
  {\enquote {\bibinfo {title} {Collective autoionization in multiply-excited
  systems: a novel ionization process observed in helium nanodroplets},}\
  }\href@noop {} {\bibfield  {journal} {\bibinfo  {journal} {Scientific
  reports}\ }\textbf {\bibinfo {volume} {4}},\ \bibinfo {pages} {3621}
  (\bibinfo {year} {2014})}\BibitemShut {NoStop}%
\bibitem [{\citenamefont {LaForge}\ \emph {et~al.}(2021)\citenamefont
  {LaForge}, \citenamefont {Michiels}, \citenamefont {Ovcharenko},
  \citenamefont {Ngai}, \citenamefont {Escart{\'\i}n}, \citenamefont {Berrah},
  \citenamefont {Callegari}, \citenamefont {Clark}, \citenamefont {Coreno},
  \citenamefont {Cucini} \emph {et~al.}}]{laforge2021ultrafast}%
  \BibitemOpen
  \bibfield  {author} {\bibinfo {author} {\bibfnamefont {A.~C.}\ \bibnamefont
  {LaForge}}, \bibinfo {author} {\bibfnamefont {R.}~\bibnamefont {Michiels}},
  \bibinfo {author} {\bibfnamefont {Y.}~\bibnamefont {Ovcharenko}}, \bibinfo
  {author} {\bibfnamefont {A.}~\bibnamefont {Ngai}}, \bibinfo {author}
  {\bibfnamefont {J.}~\bibnamefont {Escart{\'\i}n}}, \bibinfo {author}
  {\bibfnamefont {N.}~\bibnamefont {Berrah}}, \bibinfo {author} {\bibfnamefont
  {C.}~\bibnamefont {Callegari}}, \bibinfo {author} {\bibfnamefont
  {A.}~\bibnamefont {Clark}}, \bibinfo {author} {\bibfnamefont
  {M.}~\bibnamefont {Coreno}}, \bibinfo {author} {\bibfnamefont
  {R.}~\bibnamefont {Cucini}},  \emph {et~al.},\ }\bibfield  {title} {\enquote
  {\bibinfo {title} {Ultrafast resonant interatomic coulombic decay induced by
  quantum fluid dynamics},}\ }\href@noop {} {\bibfield  {journal} {\bibinfo
  {journal} {Physical Review X}\ }\textbf {\bibinfo {volume} {11}},\ \bibinfo
  {pages} {021011} (\bibinfo {year} {2021})}\BibitemShut {NoStop}%
\bibitem [{\citenamefont {Ltaief}\ \emph {et~al.}(2023)\citenamefont {Ltaief},
  \citenamefont {Sishodia}, \citenamefont {Mandal}, \citenamefont {De},
  \citenamefont {Krishnan}, \citenamefont {Medina}, \citenamefont {Pal},
  \citenamefont {Richter}, \citenamefont {Fennel},\ and\ \citenamefont
  {Mudrich}}]{ltaief2023efficient}%
  \BibitemOpen
  \bibfield  {author} {\bibinfo {author} {\bibfnamefont {L.~B.}\ \bibnamefont
  {Ltaief}}, \bibinfo {author} {\bibfnamefont {K.}~\bibnamefont {Sishodia}},
  \bibinfo {author} {\bibfnamefont {S.}~\bibnamefont {Mandal}}, \bibinfo
  {author} {\bibfnamefont {S.}~\bibnamefont {De}}, \bibinfo {author}
  {\bibfnamefont {S.}~\bibnamefont {Krishnan}}, \bibinfo {author}
  {\bibfnamefont {C.}~\bibnamefont {Medina}}, \bibinfo {author} {\bibfnamefont
  {N.}~\bibnamefont {Pal}}, \bibinfo {author} {\bibfnamefont {R.}~\bibnamefont
  {Richter}}, \bibinfo {author} {\bibfnamefont {T.}~\bibnamefont {Fennel}}, \
  and\ \bibinfo {author} {\bibfnamefont {M.}~\bibnamefont {Mudrich}},\
  }\bibfield  {title} {\enquote {\bibinfo {title} {Efficient indirect
  interatomic coulombic decay induced by photoelectron impact excitation in
  large {H}e nanodroplets},}\ }\href@noop {} {\bibfield  {journal} {\bibinfo
  {journal} {arXiv preprint arXiv:2303.14837}\ } (\bibinfo {year}
  {2023})}\BibitemShut {NoStop}%
\bibitem [{\citenamefont {Ellis}\ and\ \citenamefont
  {Yang}(2007)}]{ellis2007model}%
  \BibitemOpen
  \bibfield  {author} {\bibinfo {author} {\bibfnamefont {A.~M.}\ \bibnamefont
  {Ellis}}\ and\ \bibinfo {author} {\bibfnamefont {S.}~\bibnamefont {Yang}},\
  }\bibfield  {title} {\enquote {\bibinfo {title} {Model for the
  charge-transfer probability in helium nanodroplets following electron-impact
  ionization},}\ }\href@noop {} {\bibfield  {journal} {\bibinfo  {journal}
  {Physical Review A}\ }\textbf {\bibinfo {volume} {76}},\ \bibinfo {pages}
  {032714} (\bibinfo {year} {2007})}\BibitemShut {NoStop}%
\bibitem [{\citenamefont {Hertel}\ and\ \citenamefont
  {Hoffmann}(2011)}]{hertel2011astrid2}%
  \BibitemOpen
  \bibfield  {author} {\bibinfo {author} {\bibfnamefont {N.}~\bibnamefont
  {Hertel}}\ and\ \bibinfo {author} {\bibfnamefont {S.~V.}\ \bibnamefont
  {Hoffmann}},\ }\bibfield  {title} {\enquote {\bibinfo {title} {Astrid2: A new
  danish low-emittance sr source},}\ }\href@noop {} {\bibfield  {journal}
  {\bibinfo  {journal} {Synchrotron Radiation News}\ }\textbf {\bibinfo
  {volume} {24}},\ \bibinfo {pages} {19--23} (\bibinfo {year}
  {2011})}\BibitemShut {NoStop}%
\bibitem [{\citenamefont {Gomez}\ \emph {et~al.}(2011)\citenamefont {Gomez},
  \citenamefont {Loginov}, \citenamefont {Sliter},\ and\ \citenamefont
  {Vilesov}}]{gomez2011sizes}%
  \BibitemOpen
  \bibfield  {author} {\bibinfo {author} {\bibfnamefont {L.~F.}\ \bibnamefont
  {Gomez}}, \bibinfo {author} {\bibfnamefont {E.}~\bibnamefont {Loginov}},
  \bibinfo {author} {\bibfnamefont {R.}~\bibnamefont {Sliter}}, \ and\ \bibinfo
  {author} {\bibfnamefont {A.~F.}\ \bibnamefont {Vilesov}},\ }\bibfield
  {title} {\enquote {\bibinfo {title} {Sizes of large he droplets},}\
  }\href@noop {} {\bibfield  {journal} {\bibinfo  {journal} {J. Chem. Phys.}\
  }\textbf {\bibinfo {volume} {135}},\ \bibinfo {pages} {154201} (\bibinfo
  {year} {2011})}\BibitemShut {NoStop}%
\bibitem [{\citenamefont {Stephens}\ and\ \citenamefont
  {King}(1983)}]{stephens1983experimental}%
  \BibitemOpen
  \bibfield  {author} {\bibinfo {author} {\bibfnamefont {P.~W.}\ \bibnamefont
  {Stephens}}\ and\ \bibinfo {author} {\bibfnamefont {J.~G.}\ \bibnamefont
  {King}},\ }\bibfield  {title} {\enquote {\bibinfo {title} {Experimental
  investigation of small helium clusters: {M}agic numbers and the onset of
  condensation},}\ }\href@noop {} {\bibfield  {journal} {\bibinfo  {journal}
  {prl}\ }\textbf {\bibinfo {volume} {51}},\ \bibinfo {pages} {1538} (\bibinfo
  {year} {1983})}\BibitemShut {NoStop}%
\bibitem [{\citenamefont {Callicoatt}\ \emph
  {et~al.}(1998{\natexlab{a}})\citenamefont {Callicoatt}, \citenamefont
  {F{\"o}rde}, \citenamefont {Jung}, \citenamefont {Ruchti},\ and\
  \citenamefont {Janda}}]{callicoatt1998fragmentation}%
  \BibitemOpen
  \bibfield  {author} {\bibinfo {author} {\bibfnamefont {B.~E.}\ \bibnamefont
  {Callicoatt}}, \bibinfo {author} {\bibfnamefont {K.}~\bibnamefont
  {F{\"o}rde}}, \bibinfo {author} {\bibfnamefont {L.~F.}\ \bibnamefont {Jung}},
  \bibinfo {author} {\bibfnamefont {T.}~\bibnamefont {Ruchti}}, \ and\ \bibinfo
  {author} {\bibfnamefont {K.~C.}\ \bibnamefont {Janda}},\ }\bibfield  {title}
  {\enquote {\bibinfo {title} {Fragmentation of ionized liquid helium droplets:
  A new interpretation},}\ }\href@noop {} {\bibfield  {journal} {\bibinfo
  {journal} {J. Chem. Phys.}\ }\textbf {\bibinfo {volume} {109}},\ \bibinfo
  {pages} {10195--10200} (\bibinfo {year} {1998}{\natexlab{a}})}\BibitemShut
  {NoStop}%
\bibitem [{\citenamefont {Buchenau}\ \emph {et~al.}(1990)\citenamefont
  {Buchenau}, \citenamefont {Knuth}, \citenamefont {Northby}, \citenamefont
  {Toennies},\ and\ \citenamefont {Winkler}}]{buchenau1990mass}%
  \BibitemOpen
  \bibfield  {author} {\bibinfo {author} {\bibfnamefont {H.}~\bibnamefont
  {Buchenau}}, \bibinfo {author} {\bibfnamefont {E.}~\bibnamefont {Knuth}},
  \bibinfo {author} {\bibfnamefont {J.}~\bibnamefont {Northby}}, \bibinfo
  {author} {\bibfnamefont {J.}~\bibnamefont {Toennies}}, \ and\ \bibinfo
  {author} {\bibfnamefont {C.}~\bibnamefont {Winkler}},\ }\bibfield  {title}
  {\enquote {\bibinfo {title} {Mass spectra and time-of-flight distributions of
  helium cluster beams},}\ }\href@noop {} {\bibfield  {journal} {\bibinfo
  {journal} {J. Chem. Phys.}\ }\textbf {\bibinfo {volume} {92}},\ \bibinfo
  {pages} {6875--6889} (\bibinfo {year} {1990})}\BibitemShut {NoStop}%
\bibitem [{\citenamefont {Buchenau}, \citenamefont {Toennies},\ and\
  \citenamefont {Northby}(1991)}]{buchenau1991excitation}%
  \BibitemOpen
  \bibfield  {author} {\bibinfo {author} {\bibfnamefont {H.}~\bibnamefont
  {Buchenau}}, \bibinfo {author} {\bibfnamefont {J.}~\bibnamefont {Toennies}},
  \ and\ \bibinfo {author} {\bibfnamefont {J.}~\bibnamefont {Northby}},\
  }\bibfield  {title} {\enquote {\bibinfo {title} {Excitation and ionization of
  4{H}e clusters by electrons},}\ }\href@noop {} {\bibfield  {journal}
  {\bibinfo  {journal} {J. Chem. Phys.}\ }\textbf {\bibinfo {volume} {95}},\
  \bibinfo {pages} {8134--8148} (\bibinfo {year} {1991})}\BibitemShut {NoStop}%
\bibitem [{\citenamefont {Hellsing}\ \emph {et~al.}(1985)\citenamefont
  {Hellsing}, \citenamefont {Karlsson}, \citenamefont {Andren},\ and\
  \citenamefont {Norden}}]{hellsing1985performance}%
  \BibitemOpen
  \bibfield  {author} {\bibinfo {author} {\bibfnamefont {M.}~\bibnamefont
  {Hellsing}}, \bibinfo {author} {\bibfnamefont {L.}~\bibnamefont {Karlsson}},
  \bibinfo {author} {\bibfnamefont {H.-O.}\ \bibnamefont {Andren}}, \ and\
  \bibinfo {author} {\bibfnamefont {H.}~\bibnamefont {Norden}},\ }\bibfield
  {title} {\enquote {\bibinfo {title} {Performance of a microchannel plate ion
  detector in the energy range 3-25 ke{V}},}\ }\href@noop {} {\ \textbf
  {\bibinfo {volume} {18}},\ \bibinfo {pages} {920} (\bibinfo {year}
  {1985})}\BibitemShut {NoStop}%
\bibitem [{\citenamefont {Samson}\ and\ \citenamefont
  {Stolte}(2002)}]{samson2002precision}%
  \BibitemOpen
  \bibfield  {author} {\bibinfo {author} {\bibfnamefont {J.}~\bibnamefont
  {Samson}}\ and\ \bibinfo {author} {\bibfnamefont {W.~C.}\ \bibnamefont
  {Stolte}},\ }\bibfield  {title} {\enquote {\bibinfo {title} {Precision
  measurements of the total photoionization cross-sections of {H}e, {N}e, {A}r,
  {K}r, and {X}e},}\ }\href@noop {} {\bibfield  {journal} {\bibinfo  {journal}
  {Journal of Electron Spectroscopy and Related Phenomena}\ }\textbf {\bibinfo
  {volume} {123}},\ \bibinfo {pages} {265--276} (\bibinfo {year}
  {2002})}\BibitemShut {NoStop}%
\bibitem [{\citenamefont {Stringari}\ and\ \citenamefont
  {Treiner}(1987)}]{stringari1987systematics}%
  \BibitemOpen
  \bibfield  {author} {\bibinfo {author} {\bibfnamefont {S.}~\bibnamefont
  {Stringari}}\ and\ \bibinfo {author} {\bibfnamefont {J.}~\bibnamefont
  {Treiner}},\ }\bibfield  {title} {\enquote {\bibinfo {title} {Systematics of
  liquid helium clusters},}\ }\href@noop {} {\bibfield  {journal} {\bibinfo
  {journal} {J. Chem. Phys.}\ }\textbf {\bibinfo {volume} {87}},\ \bibinfo
  {pages} {5021--5027} (\bibinfo {year} {1987})}\BibitemShut {NoStop}%
\bibitem [{\citenamefont {Laimer}\ \emph {et~al.}(2019)\citenamefont {Laimer},
  \citenamefont {Kranabetter}, \citenamefont {Tiefenthaler}, \citenamefont
  {Albertini}, \citenamefont {Zappa}, \citenamefont {Ellis}, \citenamefont
  {Gatchell},\ and\ \citenamefont {Scheier}}]{laimer2019highly}%
  \BibitemOpen
  \bibfield  {author} {\bibinfo {author} {\bibfnamefont {F.}~\bibnamefont
  {Laimer}}, \bibinfo {author} {\bibfnamefont {L.}~\bibnamefont {Kranabetter}},
  \bibinfo {author} {\bibfnamefont {L.}~\bibnamefont {Tiefenthaler}}, \bibinfo
  {author} {\bibfnamefont {S.}~\bibnamefont {Albertini}}, \bibinfo {author}
  {\bibfnamefont {F.}~\bibnamefont {Zappa}}, \bibinfo {author} {\bibfnamefont
  {A.~M.}\ \bibnamefont {Ellis}}, \bibinfo {author} {\bibfnamefont
  {M.}~\bibnamefont {Gatchell}}, \ and\ \bibinfo {author} {\bibfnamefont
  {P.}~\bibnamefont {Scheier}},\ }\bibfield  {title} {\enquote {\bibinfo
  {title} {Highly charged droplets of superfluid helium},}\ }\href@noop {}
  {\bibfield  {journal} {\bibinfo  {journal} {Phys. Rev. Lett.}\ }\textbf
  {\bibinfo {volume} {123}},\ \bibinfo {pages} {165301} (\bibinfo {year}
  {2019})}\BibitemShut {NoStop}%
\bibitem [{\citenamefont {Peterka}\ \emph {et~al.}(2003)\citenamefont
  {Peterka}, \citenamefont {Lindinger}, \citenamefont {Poisson}, \citenamefont
  {Ahmed},\ and\ \citenamefont {Neumark}}]{peterka2003photoelectron}%
  \BibitemOpen
  \bibfield  {author} {\bibinfo {author} {\bibfnamefont {D.~S.}\ \bibnamefont
  {Peterka}}, \bibinfo {author} {\bibfnamefont {A.}~\bibnamefont {Lindinger}},
  \bibinfo {author} {\bibfnamefont {L.}~\bibnamefont {Poisson}}, \bibinfo
  {author} {\bibfnamefont {M.}~\bibnamefont {Ahmed}}, \ and\ \bibinfo {author}
  {\bibfnamefont {D.~M.}\ \bibnamefont {Neumark}},\ }\bibfield  {title}
  {\enquote {\bibinfo {title} {Photoelectron imaging of helium droplets},}\
  }\href@noop {} {\bibfield  {journal} {\bibinfo  {journal} {Phys. Rev. Lett.}\
  }\textbf {\bibinfo {volume} {91}},\ \bibinfo {pages} {043401} (\bibinfo
  {year} {2003})}\BibitemShut {NoStop}%
\bibitem [{\citenamefont {Lewerenz}, \citenamefont {Schilling},\ and\
  \citenamefont {Toennies}(1995)}]{lewerenz1995successive}%
  \BibitemOpen
  \bibfield  {author} {\bibinfo {author} {\bibfnamefont {M.}~\bibnamefont
  {Lewerenz}}, \bibinfo {author} {\bibfnamefont {B.}~\bibnamefont {Schilling}},
  \ and\ \bibinfo {author} {\bibfnamefont {J.}~\bibnamefont {Toennies}},\
  }\bibfield  {title} {\enquote {\bibinfo {title} {Successive capture and
  coagulation of atoms and molecules to small clusters in large liquid helium
  clusters},}\ }\href@noop {} {\bibfield  {journal} {\bibinfo  {journal} {J.
  Chem. Phys.}\ }\textbf {\bibinfo {volume} {102}},\ \bibinfo {pages}
  {8191--8207} (\bibinfo {year} {1995})}\BibitemShut {NoStop}%
\bibitem [{\citenamefont {Hessj}, \citenamefont {Larsen},\ and\ \citenamefont
  {Scheidemann}(1999)}]{hessj1999measurement}%
  \BibitemOpen
  \bibfield  {author} {\bibinfo {author} {\bibfnamefont {H.}~\bibnamefont
  {Hessj}}, \bibinfo {author} {\bibfnamefont {D.~S.}\ \bibnamefont {Larsen}}, \
  and\ \bibinfo {author} {\bibfnamefont {A.~A.}\ \bibnamefont {Scheidemann}},\
  }\bibfield  {title} {\enquote {\bibinfo {title} {Measurement of pick-up
  cross-sections of 4{H}e clusters: {P}olar versus non-polar molecules},}\
  }\href@noop {} {\bibfield  {journal} {\bibinfo  {journal} {Philosophical
  Magazine B}\ }\textbf {\bibinfo {volume} {79}},\ \bibinfo {pages}
  {1437--1444} (\bibinfo {year} {1999})}\BibitemShut {NoStop}%
\bibitem [{\citenamefont {B{\"u}nermann}\ and\ \citenamefont
  {Stienkemeier}(2011)}]{bunermann2011modeling}%
  \BibitemOpen
  \bibfield  {author} {\bibinfo {author} {\bibfnamefont {O.}~\bibnamefont
  {B{\"u}nermann}}\ and\ \bibinfo {author} {\bibfnamefont {F.}~\bibnamefont
  {Stienkemeier}},\ }\bibfield  {title} {\enquote {\bibinfo {title} {Modeling
  the formation of alkali clusters attached to helium nanodroplets and the
  abundance of high-spin states},}\ }\href@noop {} {\bibfield  {journal}
  {\bibinfo  {journal} {The European Physical Journal D}\ }\textbf {\bibinfo
  {volume} {61}},\ \bibinfo {pages} {645--655} (\bibinfo {year}
  {2011})}\BibitemShut {NoStop}%
\bibitem [{\citenamefont {Brink}\ and\ \citenamefont
  {Stringari}(1990)}]{brink1990density}%
  \BibitemOpen
  \bibfield  {author} {\bibinfo {author} {\bibfnamefont {D.}~\bibnamefont
  {Brink}}\ and\ \bibinfo {author} {\bibfnamefont {S.}~\bibnamefont
  {Stringari}},\ }\bibfield  {title} {\enquote {\bibinfo {title} {Density of
  states and evaporation rate of helium clusters},}\ }\href@noop {} {\bibfield
  {journal} {\bibinfo  {journal} {Zeitschrift f{\"u}r Physik D Atoms, Molecules
  and Clusters}\ }\textbf {\bibinfo {volume} {15}},\ \bibinfo {pages}
  {257--263} (\bibinfo {year} {1990})}\BibitemShut {NoStop}%
\bibitem [{\citenamefont {Lehmann}\ and\ \citenamefont
  {Dokter}(2004)}]{lehmann2004evaporative}%
  \BibitemOpen
  \bibfield  {author} {\bibinfo {author} {\bibfnamefont {K.~K.}\ \bibnamefont
  {Lehmann}}\ and\ \bibinfo {author} {\bibfnamefont {A.~M.}\ \bibnamefont
  {Dokter}},\ }\bibfield  {title} {\enquote {\bibinfo {title} {Evaporative
  cooling of helium nanodroplets with angular momentum conservation},}\
  }\href@noop {} {\bibfield  {journal} {\bibinfo  {journal} {Phys. Rev. Lett.}\
  }\textbf {\bibinfo {volume} {92}},\ \bibinfo {pages} {173401} (\bibinfo
  {year} {2004})}\BibitemShut {NoStop}%
\bibitem [{\citenamefont {Callicoatt}\ \emph
  {et~al.}(1998{\natexlab{b}})\citenamefont {Callicoatt}, \citenamefont
  {F{\"o}rde}, \citenamefont {Ruchti}, \citenamefont {Jung}, \citenamefont
  {Janda},\ and\ \citenamefont {Halberstadt}}]{callicoatt1998capture}%
  \BibitemOpen
  \bibfield  {author} {\bibinfo {author} {\bibfnamefont {B.~E.}\ \bibnamefont
  {Callicoatt}}, \bibinfo {author} {\bibfnamefont {K.}~\bibnamefont
  {F{\"o}rde}}, \bibinfo {author} {\bibfnamefont {T.}~\bibnamefont {Ruchti}},
  \bibinfo {author} {\bibfnamefont {L.}~\bibnamefont {Jung}}, \bibinfo {author}
  {\bibfnamefont {K.~C.}\ \bibnamefont {Janda}}, \ and\ \bibinfo {author}
  {\bibfnamefont {N.}~\bibnamefont {Halberstadt}},\ }\bibfield  {title}
  {\enquote {\bibinfo {title} {Capture and ionization of argon within liquid
  helium droplets},}\ }\href@noop {} {\bibfield  {journal} {\bibinfo  {journal}
  {J. Chem. Phys.}\ }\textbf {\bibinfo {volume} {108}},\ \bibinfo {pages}
  {9371--9382} (\bibinfo {year} {1998}{\natexlab{b}})}\BibitemShut {NoStop}%
\bibitem [{\citenamefont {Halberstadt}\ and\ \citenamefont
  {Janda}(1998)}]{halberstadt1998resonant}%
  \BibitemOpen
  \bibfield  {author} {\bibinfo {author} {\bibfnamefont {N.}~\bibnamefont
  {Halberstadt}}\ and\ \bibinfo {author} {\bibfnamefont {K.~C.}\ \bibnamefont
  {Janda}},\ }\bibfield  {title} {\enquote {\bibinfo {title} {The resonant
  charge hopping rate in positively charged helium clusters},}\ }\href@noop {}
  {\bibfield  {journal} {\bibinfo  {journal} {Chem. Phys. Lett.}\ }\textbf
  {\bibinfo {volume} {282}},\ \bibinfo {pages} {409--412} (\bibinfo {year}
  {1998})}\BibitemShut {NoStop}%
\bibitem [{\citenamefont {Lewis}\ \emph {et~al.}(2005)\citenamefont {Lewis},
  \citenamefont {Lindsay}, \citenamefont {Bemish},\ and\ \citenamefont
  {Miller}}]{lewis2005probing}%
  \BibitemOpen
  \bibfield  {author} {\bibinfo {author} {\bibfnamefont {W.~K.}\ \bibnamefont
  {Lewis}}, \bibinfo {author} {\bibfnamefont {C.~M.}\ \bibnamefont {Lindsay}},
  \bibinfo {author} {\bibfnamefont {R.~J.}\ \bibnamefont {Bemish}}, \ and\
  \bibinfo {author} {\bibfnamefont {R.~E.}\ \bibnamefont {Miller}},\ }\bibfield
   {title} {\enquote {\bibinfo {title} {Probing charge-transfer processes in
  helium nanodroplets by optically selected mass spectrometry ({OSMS}):
  {C}harge steering by long-range interactions},}\ }\href@noop {} {\bibfield
  {journal} {\bibinfo  {journal} {Journal of the American Chemical Society}\
  }\textbf {\bibinfo {volume} {127}},\ \bibinfo {pages} {7235--7242} (\bibinfo
  {year} {2005})}\BibitemShut {NoStop}%
\bibitem [{\citenamefont {Yousif}\ \emph {et~al.}(1987)\citenamefont {Yousif},
  \citenamefont {Lindsay}, \citenamefont {Simpson},\ and\ \citenamefont
  {Latimer}}]{yousif1987dissociative}%
  \BibitemOpen
  \bibfield  {author} {\bibinfo {author} {\bibfnamefont {F.}~\bibnamefont
  {Yousif}}, \bibinfo {author} {\bibfnamefont {B.}~\bibnamefont {Lindsay}},
  \bibinfo {author} {\bibfnamefont {F.}~\bibnamefont {Simpson}}, \ and\
  \bibinfo {author} {\bibfnamefont {C.}~\bibnamefont {Latimer}},\ }\bibfield
  {title} {\enquote {\bibinfo {title} {Dissociative ionisation and charge
  transfer in {H}e+-{O}2 collisions},}\ }\href@noop {} {\bibfield  {journal}
  {\bibinfo  {journal} {J. Phys. B: At., Mol. Opt. Phys.}\ }\textbf {\bibinfo
  {volume} {20}},\ \bibinfo {pages} {5079} (\bibinfo {year}
  {1987})}\BibitemShut {NoStop}%
\bibitem [{\citenamefont {Callicoatt}\ \emph {et~al.}(1996)\citenamefont
  {Callicoatt}, \citenamefont {Mar}, \citenamefont {Apkarian},\ and\
  \citenamefont {Janda}}]{callicoatt1996charge}%
  \BibitemOpen
  \bibfield  {author} {\bibinfo {author} {\bibfnamefont {B.~E.}\ \bibnamefont
  {Callicoatt}}, \bibinfo {author} {\bibfnamefont {D.~D.}\ \bibnamefont {Mar}},
  \bibinfo {author} {\bibfnamefont {V.}~\bibnamefont {Apkarian}}, \ and\
  \bibinfo {author} {\bibfnamefont {K.~C.}\ \bibnamefont {Janda}},\ }\bibfield
  {title} {\enquote {\bibinfo {title} {Charge transfer within {H}e clusters},}\
  }\href@noop {} {\bibfield  {journal} {\bibinfo  {journal} {J. Chem. Phys.}\
  }\textbf {\bibinfo {volume} {105}},\ \bibinfo {pages} {7872--7875} (\bibinfo
  {year} {1996})}\BibitemShut {NoStop}%
\bibitem [{\citenamefont {Ruchti}\ \emph {et~al.}(1998)\citenamefont {Ruchti},
  \citenamefont {F{\"o}rde}, \citenamefont {Callicoatt}, \citenamefont
  {Ludwigs},\ and\ \citenamefont {Janda}}]{ruchti1998charge}%
  \BibitemOpen
  \bibfield  {author} {\bibinfo {author} {\bibfnamefont {T.}~\bibnamefont
  {Ruchti}}, \bibinfo {author} {\bibfnamefont {K.}~\bibnamefont {F{\"o}rde}},
  \bibinfo {author} {\bibfnamefont {B.~E.}\ \bibnamefont {Callicoatt}},
  \bibinfo {author} {\bibfnamefont {H.}~\bibnamefont {Ludwigs}}, \ and\
  \bibinfo {author} {\bibfnamefont {K.~C.}\ \bibnamefont {Janda}},\ }\bibfield
  {title} {\enquote {\bibinfo {title} {Charge transfer and fragmentation of
  liquid helium clusters that contain one or more neon atoms},}\ }\href@noop {}
  {\bibfield  {journal} {\bibinfo  {journal} {J. Chem. Phys.}\ }\textbf
  {\bibinfo {volume} {109}},\ \bibinfo {pages} {10679--10687} (\bibinfo {year}
  {1998})}\BibitemShut {NoStop}%
\bibitem [{\citenamefont {W{\"o}rmer}\ \emph {et~al.}(1989)\citenamefont
  {W{\"o}rmer}, \citenamefont {Guzielski}, \citenamefont {Stapelfeldt},\ and\
  \citenamefont {M{\"o}ller}}]{wormer1989fluorescence}%
  \BibitemOpen
  \bibfield  {author} {\bibinfo {author} {\bibfnamefont {J.}~\bibnamefont
  {W{\"o}rmer}}, \bibinfo {author} {\bibfnamefont {V.}~\bibnamefont
  {Guzielski}}, \bibinfo {author} {\bibfnamefont {J.}~\bibnamefont
  {Stapelfeldt}}, \ and\ \bibinfo {author} {\bibfnamefont {T.}~\bibnamefont
  {M{\"o}ller}},\ }\bibfield  {title} {\enquote {\bibinfo {title} {Fluorescence
  excitation spectroscopy of xenon clusters in the {VUV}},}\ }\href@noop {}
  {\bibfield  {journal} {\bibinfo  {journal} {Chem. Phys. Lett.}\ }\textbf
  {\bibinfo {volume} {159}},\ \bibinfo {pages} {321--326} (\bibinfo {year}
  {1989})}\BibitemShut {NoStop}%
\bibitem [{\citenamefont {Stapelfeldt}, \citenamefont {W{\"o}rmer},\ and\
  \citenamefont {M{\"o}ller}(1989)}]{stapelfeldt1989evolution}%
  \BibitemOpen
  \bibfield  {author} {\bibinfo {author} {\bibfnamefont {J.}~\bibnamefont
  {Stapelfeldt}}, \bibinfo {author} {\bibfnamefont {J.}~\bibnamefont
  {W{\"o}rmer}}, \ and\ \bibinfo {author} {\bibfnamefont {T.}~\bibnamefont
  {M{\"o}ller}},\ }\bibfield  {title} {\enquote {\bibinfo {title} {Evolution of
  electronic energy levels in krypton clusters from the atom to the solid},}\
  }\href@noop {} {\bibfield  {journal} {\bibinfo  {journal} {Phys. Rev. Lett.}\
  }\textbf {\bibinfo {volume} {62}},\ \bibinfo {pages} {98} (\bibinfo {year}
  {1989})}\BibitemShut {NoStop}%
\bibitem [{\citenamefont {Closser}, \citenamefont {Gessner},\ and\
  \citenamefont {Head-Gordon}(2014)}]{closser2014simulations}%
  \BibitemOpen
  \bibfield  {author} {\bibinfo {author} {\bibfnamefont {K.~D.}\ \bibnamefont
  {Closser}}, \bibinfo {author} {\bibfnamefont {O.}~\bibnamefont {Gessner}}, \
  and\ \bibinfo {author} {\bibfnamefont {M.}~\bibnamefont {Head-Gordon}},\
  }\bibfield  {title} {\enquote {\bibinfo {title} {Simulations of the
  dissociation of small helium clusters with ab initio molecular dynamics in
  electronically excited states},}\ }\href@noop {} {\bibfield  {journal}
  {\bibinfo  {journal} {J. Chem. Phys.}\ }\textbf {\bibinfo {volume} {140}},\
  \bibinfo {pages} {134306} (\bibinfo {year} {2014})}\BibitemShut {NoStop}%
\bibitem [{\citenamefont {Leal}\ \emph {et~al.}(2014)\citenamefont {Leal},
  \citenamefont {Mateo}, \citenamefont {Hernando}, \citenamefont {Pi},
  \citenamefont {Barranco}, \citenamefont {Ponti}, \citenamefont {Cargnoni},\
  and\ \citenamefont {Drabbels}}]{leal2014picosecond}%
  \BibitemOpen
  \bibfield  {author} {\bibinfo {author} {\bibfnamefont {A.}~\bibnamefont
  {Leal}}, \bibinfo {author} {\bibfnamefont {D.}~\bibnamefont {Mateo}},
  \bibinfo {author} {\bibfnamefont {A.}~\bibnamefont {Hernando}}, \bibinfo
  {author} {\bibfnamefont {M.}~\bibnamefont {Pi}}, \bibinfo {author}
  {\bibfnamefont {M.}~\bibnamefont {Barranco}}, \bibinfo {author}
  {\bibfnamefont {A.}~\bibnamefont {Ponti}}, \bibinfo {author} {\bibfnamefont
  {F.}~\bibnamefont {Cargnoni}}, \ and\ \bibinfo {author} {\bibfnamefont
  {M.}~\bibnamefont {Drabbels}},\ }\bibfield  {title} {\enquote {\bibinfo
  {title} {Picosecond solvation dynamics of alkali cations in superfluid {H}e 4
  nanodroplets},}\ }\href@noop {} {\bibfield  {journal} {\bibinfo  {journal}
  {Physical Review B}\ }\textbf {\bibinfo {volume} {90}},\ \bibinfo {pages}
  {224518} (\bibinfo {year} {2014})}\BibitemShut {NoStop}%
\bibitem [{\citenamefont {Ancilotto}, \citenamefont {DeToffol},\ and\
  \citenamefont {Toigo}(1995)}]{ancilotto1995sodium}%
  \BibitemOpen
  \bibfield  {author} {\bibinfo {author} {\bibfnamefont {F.}~\bibnamefont
  {Ancilotto}}, \bibinfo {author} {\bibfnamefont {G.}~\bibnamefont {DeToffol}},
  \ and\ \bibinfo {author} {\bibfnamefont {F.}~\bibnamefont {Toigo}},\
  }\bibfield  {title} {\enquote {\bibinfo {title} {Sodium dimers on the surface
  of liquid {H}e4},}\ }\href@noop {} {\bibfield  {journal} {\bibinfo  {journal}
  {Physical Review B}\ }\textbf {\bibinfo {volume} {52}},\ \bibinfo {pages}
  {16125} (\bibinfo {year} {1995})}\BibitemShut {NoStop}%
\bibitem [{\citenamefont {An~der Lan}\ \emph {et~al.}(2011)\citenamefont
  {An~der Lan}, \citenamefont {Bartl}, \citenamefont {Leidlmair}, \citenamefont
  {Sch{\"o}bel}, \citenamefont {Jochum}, \citenamefont {Denifl}, \citenamefont
  {M{\"a}rk}, \citenamefont {Ellis},\ and\ \citenamefont
  {Scheier}}]{an2011submersion}%
  \BibitemOpen
  \bibfield  {author} {\bibinfo {author} {\bibfnamefont {L.}~\bibnamefont
  {An~der Lan}}, \bibinfo {author} {\bibfnamefont {P.}~\bibnamefont {Bartl}},
  \bibinfo {author} {\bibfnamefont {C.}~\bibnamefont {Leidlmair}}, \bibinfo
  {author} {\bibfnamefont {H.}~\bibnamefont {Sch{\"o}bel}}, \bibinfo {author}
  {\bibfnamefont {R.}~\bibnamefont {Jochum}}, \bibinfo {author} {\bibfnamefont
  {S.}~\bibnamefont {Denifl}}, \bibinfo {author} {\bibfnamefont {T.~D.}\
  \bibnamefont {M{\"a}rk}}, \bibinfo {author} {\bibfnamefont {A.~M.}\
  \bibnamefont {Ellis}}, \ and\ \bibinfo {author} {\bibfnamefont
  {P.}~\bibnamefont {Scheier}},\ }\bibfield  {title} {\enquote {\bibinfo
  {title} {The submersion of sodium clusters in helium nanodroplets:
  Identification of the surface to interior transition},}\ }\href@noop {}
  {\bibfield  {journal} {\bibinfo  {journal} {jcp}\ }\textbf {\bibinfo {volume}
  {135}},\ \bibinfo {pages} {044309} (\bibinfo {year} {2011})}\BibitemShut
  {NoStop}%
\bibitem [{\citenamefont {Messner}\ \emph {et~al.}(2022)\citenamefont
  {Messner}, \citenamefont {Di~Vora}, \citenamefont {Ernst},\ and\
  \citenamefont {Lackner}}]{messner2022photoabsorption}%
  \BibitemOpen
  \bibfield  {author} {\bibinfo {author} {\bibfnamefont {R.}~\bibnamefont
  {Messner}}, \bibinfo {author} {\bibfnamefont {R.}~\bibnamefont {Di~Vora}},
  \bibinfo {author} {\bibfnamefont {W.~E.}\ \bibnamefont {Ernst}}, \ and\
  \bibinfo {author} {\bibfnamefont {F.}~\bibnamefont {Lackner}},\ }\bibfield
  {title} {\enquote {\bibinfo {title} {Photoabsorption of potassium clusters
  isolated in helium droplets: {F}rom discrete electronic transitions to
  collective resonances},}\ }\href@noop {} {\bibfield  {journal} {\bibinfo
  {journal} {Physical Review Research}\ }\textbf {\bibinfo {volume} {4}},\
  \bibinfo {pages} {023148} (\bibinfo {year} {2022})}\BibitemShut {NoStop}%
\bibitem [{\citenamefont {M{\"u}ller}\ \emph {et~al.}(2009)\citenamefont
  {M{\"u}ller}, \citenamefont {Krapf}, \citenamefont {Koslowski}, \citenamefont
  {Mudrich},\ and\ \citenamefont {Stienkemeier}}]{muller2009cold}%
  \BibitemOpen
  \bibfield  {author} {\bibinfo {author} {\bibfnamefont {S.}~\bibnamefont
  {M{\"u}ller}}, \bibinfo {author} {\bibfnamefont {S.}~\bibnamefont {Krapf}},
  \bibinfo {author} {\bibfnamefont {T.}~\bibnamefont {Koslowski}}, \bibinfo
  {author} {\bibfnamefont {M.}~\bibnamefont {Mudrich}}, \ and\ \bibinfo
  {author} {\bibfnamefont {F.}~\bibnamefont {Stienkemeier}},\ }\bibfield
  {title} {\enquote {\bibinfo {title} {Cold reactions of alkali-metal and water
  clusters inside helium nanodroplets},}\ }\href@noop {} {\bibfield  {journal}
  {\bibinfo  {journal} {Phys. Rev. Lett.}\ }\textbf {\bibinfo {volume} {102}},\
  \bibinfo {pages} {183401} (\bibinfo {year} {2009})}\BibitemShut {NoStop}%
\bibitem [{\citenamefont {Schulz}\ \emph {et~al.}(2004)\citenamefont {Schulz},
  \citenamefont {Claas}, \citenamefont {Schumacher},\ and\ \citenamefont
  {Stienkemeier}}]{schulz2004formation}%
  \BibitemOpen
  \bibfield  {author} {\bibinfo {author} {\bibfnamefont {C.}~\bibnamefont
  {Schulz}}, \bibinfo {author} {\bibfnamefont {P.}~\bibnamefont {Claas}},
  \bibinfo {author} {\bibfnamefont {D.}~\bibnamefont {Schumacher}}, \ and\
  \bibinfo {author} {\bibfnamefont {F.}~\bibnamefont {Stienkemeier}},\
  }\bibfield  {title} {\enquote {\bibinfo {title} {Formation and stability of
  high-spin alkali clusters},}\ }\href@noop {} {\bibfield  {journal} {\bibinfo
  {journal} {Phys. Rev. Lett.}\ }\textbf {\bibinfo {volume} {92}},\ \bibinfo
  {pages} {013401} (\bibinfo {year} {2004})}\BibitemShut {NoStop}%
\bibitem [{\citenamefont {Joppien}, \citenamefont {Karnbach},\ and\
  \citenamefont {M{\"o}ller}(1993)}]{joppien1993electronic}%
  \BibitemOpen
  \bibfield  {author} {\bibinfo {author} {\bibfnamefont {M.}~\bibnamefont
  {Joppien}}, \bibinfo {author} {\bibfnamefont {R.}~\bibnamefont {Karnbach}}, \
  and\ \bibinfo {author} {\bibfnamefont {T.}~\bibnamefont {M{\"o}ller}},\
  }\bibfield  {title} {\enquote {\bibinfo {title} {Electronic excitations in
  liquid helium: The evolution from small clusters to large droplets},}\
  }\href@noop {} {\bibfield  {journal} {\bibinfo  {journal} {Phys. Rev. Lett.}\
  }\textbf {\bibinfo {volume} {71}},\ \bibinfo {pages} {2654} (\bibinfo {year}
  {1993})}\BibitemShut {NoStop}%
\bibitem [{\citenamefont {Closser}\ and\ \citenamefont
  {Head-Gordon}(2010)}]{closser2010ab}%
  \BibitemOpen
  \bibfield  {author} {\bibinfo {author} {\bibfnamefont {K.~D.}\ \bibnamefont
  {Closser}}\ and\ \bibinfo {author} {\bibfnamefont {M.}~\bibnamefont
  {Head-Gordon}},\ }\bibfield  {title} {\enquote {\bibinfo {title} {Ab initio
  calculations on the electronically excited states of small helium
  clusters},}\ }\href@noop {} {\ \textbf {\bibinfo {volume} {114}},\ \bibinfo
  {pages} {8023--8032} (\bibinfo {year} {2010})}\BibitemShut {NoStop}%
\bibitem [{\citenamefont {Haddad}\ and\ \citenamefont
  {Samson}(1986)}]{haddad1986total}%
  \BibitemOpen
  \bibfield  {author} {\bibinfo {author} {\bibfnamefont {G.}~\bibnamefont
  {Haddad}}\ and\ \bibinfo {author} {\bibfnamefont {J.~A.}\ \bibnamefont
  {Samson}},\ }\bibfield  {title} {\enquote {\bibinfo {title} {Total absorption
  and photoionization cross sections of water vapor between 100 and 1000
  {Å}},}\ }\href@noop {} {\bibfield  {journal} {\bibinfo  {journal} {J. Chem.
  Phys.}\ }\textbf {\bibinfo {volume} {84}},\ \bibinfo {pages} {6623--6626}
  (\bibinfo {year} {1986})}\BibitemShut {NoStop}%
\bibitem [{\citenamefont {Frenking}\ \emph {et~al.}(1989)\citenamefont
  {Frenking}, \citenamefont {Koch}, \citenamefont {Cremer}, \citenamefont
  {Gauss},\ and\ \citenamefont {Liebman}}]{frenking1989helium}%
  \BibitemOpen
  \bibfield  {author} {\bibinfo {author} {\bibfnamefont {G.}~\bibnamefont
  {Frenking}}, \bibinfo {author} {\bibfnamefont {W.}~\bibnamefont {Koch}},
  \bibinfo {author} {\bibfnamefont {D.}~\bibnamefont {Cremer}}, \bibinfo
  {author} {\bibfnamefont {J.}~\bibnamefont {Gauss}}, \ and\ \bibinfo {author}
  {\bibfnamefont {J.~F.}\ \bibnamefont {Liebman}},\ }\bibfield  {title}
  {\enquote {\bibinfo {title} {Helium bonding in singly and doubly charged
  first-row diatomic cations {H}e{X}n+ ({X}= {L}i-{N}e; n= 1, 2)},}\
  }\href@noop {} {\bibfield  {journal} {\bibinfo  {journal} {The Journal of
  Physical Chemistry}\ }\textbf {\bibinfo {volume} {93}},\ \bibinfo {pages}
  {3397--3410} (\bibinfo {year} {1989})}\BibitemShut {NoStop}%
\end{thebibliography}%

\end{document}